\documentclass[english,aps,superscriptaddress,floatfix, nofootinbib, notitlepage, longbibliography]{revtex4-2}
\usepackage[utf8]{inputenc}
\usepackage{color}
\usepackage{bm}
\usepackage{xcolor}
\usepackage{amsmath}
\usepackage{amssymb}
\usepackage{graphicx}
\usepackage[sort&compress]{natbib}
\usepackage{braket}
\usepackage{psfrag}
\usepackage{tikz}
\usepackage[normalem]{ulem}
\usepackage{qcircuit}
\usepackage{ulem}
\usepackage{multirow}
\usepackage[ruled,linesnumbered,vlined]{algorithm2e}
\usepackage{hyperref}
\hypersetup{colorlinks=true, citecolor=blue, urlcolor=blue, linkcolor=blue}
\usepackage[caption=false]{subfig}
\usepackage{esint}
\usepackage{kbordermatrix}
\allowdisplaybreaks

\newcommand{\var}{VaR}
\newcommand{\cvar}{CVaR}
\newcommand{\varvalue}{V_{\alpha}}
\newcommand{\cvarvalue}{C_{\alpha}}
\newcommand{\AS}{\mathcal{A}_{\mathcal{S}}}
\newcommand{\nqae}{N_{\mu}^{\textrm{QAE}}}
\newcommand{\nqsp}{N_{\mu}^{\textrm{QSP}}}

\newcommand{\identity}{\mathbb{I}}
\newcommand{\bigO}[1]{\mathcal{O}\left( #1 \right)}
\newcommand{\epsqae}{\epsilon_{{A}}}

\begin{document}

\title{Quantum Risk Analysis of Financial Derivatives}

\author{Nikitas Stamatopoulos}
\affiliation{Goldman Sachs, New York, NY}

\author{B. David Clader}
\thanks{Current affiliation: BQP Advisors LLC}
\affiliation{Goldman Sachs, New York, NY}

\author{Stefan Woerner}
\affiliation{IBM Quantum, IBM Research Europe -- Zurich}

\author{William J. Zeng}
\thanks{Current affiliation: Quantonation}
\affiliation{Goldman Sachs, New York, NY}


\begin{abstract}
	We introduce two quantum algorithms to compute the Value at Risk (\var{}) and Conditional Value at Risk (\cvar{}) of financial derivatives using quantum computers: the first by applying existing ideas from quantum risk analysis to derivative pricing, and the second based on a novel approach using Quantum Signal Processing (QSP).
	Previous work in the literature has shown that quantum advantage is possible in the context of individual derivative pricing and that advantage can be leveraged in a straightforward manner in the estimation of the \var{} and \cvar{}.
	The algorithms we introduce in this work aim to provide an additional advantage by encoding the derivative price over multiple market scenarios in superposition and computing the desired values by applying appropriate transformations to the quantum system.
	We perform complexity and error analysis of both algorithms, and show that while the two algorithms have the same asymptotic scaling the QSP-based approach requires significantly fewer quantum resources for the same target accuracy.
	Additionally, by numerically simulating both quantum and classical \var{} algorithms, we demonstrate that the quantum algorithm can extract additional advantage from a quantum computer compared to individual derivative pricing.
	Specifically, we show that under certain conditions \var{} estimation can lower the latest published estimates of the logical clock rate required for quantum advantage in derivative pricing by up to $\sim 30$x.
	In light of these results, we are encouraged that our formulation of derivative pricing in the QSP framework may be further leveraged for quantum advantage in other relevant financial applications, and that quantum computers could be harnessed more efficiently by considering problems in the financial sector at a higher level.
\end{abstract}

\maketitle

\section{Introduction}
Ever since it was shown that quantum speedups were possible for Monte Carlo methods \cite{montanaro2015quantum}, various methods have been proposed to apply the potential benefits of quantum computing to the pricing of financial derivatives \cite{rebentrost2018quantum, Woerner_2019, Stamatopoulos_2020, Herbert_2022}.
While these quantum methods allow for a quadratic speedup over state of the art classical approaches, the quantum resources required for practical advantage are significant not just in terms of circuit depth and number of qubits, but also in terms of the logical clock rate needed to match the performance of modern classical computers \cite{chakrabarti2021threshold}.
On the other hand, the application of quantum gradient methods utilizing the quantum oracles used for pricing \cite{Stamatopoulos_2022}, showed that the logical clock rate required for advantage in the task of computing the market risk of financial derivatives might be lower than that of derivative pricing itself.
As such, considering the quantum oracles used in derivative pricing as components of higher-level algorithms which compute other quantities of interest of financial derivatives could lower certain quantum resources required for practical advantage.

Value at Risk (\var{}) is a metric that estimates the maximum possible financial depreciation of an asset (or collection of assets) within a specific time frame at a certain confidence level \cite{Markowitz1952, Roy1952}.
A related quantity, the Conditional Value at Risk (\cvar{}) measures the expected depreciation of an asset should losses exceed the Value at Risk, providing a risk metric for the tail of the loss distribution.
Financial firms make extensive use of the \var{} and \cvar{} metrics for their derivative holdings, aimed at assessing the financial health of their balance sheet as well as to set trading limits in order to minimize potential exposure to adverse market conditions.
When individual derivative contracts are priced using Monte Carlo simulation, the computation of the \var{} of portfolios of derivatives requires pricing each contract multiple times over different possible market states and then estimating the \var{} from the resulting prices.
For $N$ such market states, the complexity of estimating the \var{} thus scales as $\bigO{N/\epsilon^2}$, where $\epsilon$ is the accuracy with which each contract is priced.
The same approach can be used to estimate the \var{} by pricing of each derivative on a quantum computer and the quadratic speedup in the individual derivative pricing would carry through to the overall complexity, improving it to $\bigO{N/\epsilon}$.
In Ref.~\cite{Woerner_2019, egger2019credit}, a quantum method based on Quantum Amplitude Estimation (QAE) was introduced to estimate the \var{} in cases where potential future losses of a financial asset can be modeled explicitly.
While this method was shown to also provide a quadratic advantage compared to classical \var{} estimation using Monte Carlo, it cannot be readily applied to cases where the financial assets in question are derivative contracts which are individually priced using sampling methods, and estimating the potential losses requires repeated pricing of the contracts under different market conditions.

In this article, we introduce two quantum algorithms to estimate the \var{} and \cvar{} of derivative portfolios.
The first is an extension of the QAE-based method of Ref.~\cite{Woerner_2019, egger2019credit}.
The second is a novel algorithm based on the Quantum Signal Processing (QSP) framework \cite{Low2017optimal, gilyen2019quantum, low2019hamiltonian, Kikuchi_2023}, which enables nearly arbitrary polynomial transformations to quantum (sub)systems with minimal overheads.
Various quantum algorithms have been formulated in this framework \cite{martyn2021grand}, and in most instances the QSP formulation has resulted in lower resource requirements for the algorithms \cite{martyn2021efficient, Lin_2020, Rall_2023}, including in quantum derivative pricing \cite{stamatopoulos2023derivative}.
Moreover, it has been shown that the quantum oracles employed in derivative pricing naturally represent block-encodings of derivative prices across different input parameters \cite{gilyen2019optimizing, Stamatopoulos_2022}.
Because the QSP framework relies on efficiently constructing block-encodings of quantities of interest, we hence examine how QSP can be harnessed for problems around derivative pricing.

Leveraging the QSP framework and derivative pricing oracles, we show how to apply coherent transformations to the price of a financial derivative across multiple inputs in superposition, and create tailored transformations that allow us to estimate the \var{} and \cvar{} of derivative portfolios.
We perform complexity and error analysis of the two quantum algorithms introduced and compare their performance through explicit construction of the corresponding quantum circuits and numerical simulations.
Finally, we benchmark the QSP-based algorithm against classical \var{} estimation methods and determine that quantum advantage is possible if suitable techniques to prepare and sample from the required probability distributions are available.

In Sec.~\ref{sec:derivative_var} we formally introduce the problem of estimating the \var{} and \cvar{} of derivatives and give an overview of the QAE-based \var{} estimation method from Ref.~\cite{Woerner_2019, egger2019credit}.
In Sec.~\ref{sec:qae_var} we generalize this method to compute the \var{} of a portfolio of derivatives.
In Sec.~\ref{sec:qsp_var} we give an overview of Quantum Signal Processing, show how it can be used to perform transformations to appropriately encoded derivative prices and introduce the QSP-based \var{} and \cvar{} estimation algorithm.
In Sec.~\ref{sec:compare_methods} we compare the performance of the QAE and QSP methods in terms of the quantum resources required for a target \var{} accuracy, and in Sec.~\ref{sec:quantum_advantage} we explore the possibility of quantum advantage in estimating the \var{} of derivative portfolios.
We discuss our results and next steps in Sec.~\ref{sec:discussion}.

\section{Value at Risk}
\label{sec:derivative_var}

The \var{} computation of derivative portfolios requires modeling the possible future behavior of the market parameters underlying the derivatives.
While the pricing of derivatives relies on the value of underlying parameters today (e.g. stock prices, interest rates, volatilities, etc.), which are available (or can be estimated) from public markets, computing the \var{} of derivatives requires possible values of these parameters in the future.
Two common modeling methods for \var{} computation consist of a) using Monte Carlo simulation of the market parameters according to an appropriately defined underlying stochastic process, and b) drawing historical samples of (correlated) values for all relevant market parameters.
Each such modeled market state is also called a \emph{scenario} and can be defined as a collection of \emph{tweaks} to today's market parameters.
To price a derivative under a scenario, the tweaks are applied to the initial market state and the derivative is priced as if the tweaked market state was today's market.
Examples of scenarios defined this way could be
\begin{enumerate}
	\item Tweak asset X's price up by 5\%
	\item Tweak asset X's price up by 5\% and asset Y's price down by 4\%
	\item Tweak asset X's price up by 5\%, all points on its volatility surface up by 10\% and its correlation to asset Y by 20\%
\end{enumerate}
In both Monte Carlo and historical \var{} methods, the value of the derivative portfolio is computed under every scenario and from the resulting distribution of gains/losses, maximum levels of potential loss can be derived at different confidence levels.

More concretely, in order to compute the \var{} of a derivative portfolio at a confidence level $\alpha \in [0,1]$ (where $\alpha \ge 0.9$ is typically used), we price the portfolio under a set of $N$ scenarios $[s_1, s_2, \cdots, s_N]$ to get portfolio values $\mathcal{V} = \{V(s_i)\}$, and find the smallest portfolio value in the set such that smaller values have a probability of occurrence greater than or equal to $1-\alpha$

\begin{equation}
	\label{eqn:var_definition}
	\varvalue = \textrm{inf } \{x \; | \; \sum_{\{V \in \mathcal{V} | V \le x\}}\mathbb{P}[V] \ge 1-\alpha\},
\end{equation}
where $\mathbb{P}[V]$ denotes the probability that the portfolio has value $V$ under the evaluated scenarios.
The VaR of the portfolio at the $\alpha$ confidence level is then defined as $\textrm{VaR}_{\alpha}[V] \equiv V_0 - \varvalue$, where $V_0$ denotes the current value of the portfolio.
This approach for computing the \var{} of a portfolio can be used in both the Monte Carlo and historical methods, the only difference being the way probabilities are assigned to possible future market realizations.

The \cvar{} of a derivative portfolio measures the expected loss of the portfolio beyond the \var{} value and can be estimated using the same set of portfolio values $\mathcal{V}$ calculated for the \var{}, by computing the quantity

\begin{equation}
	\label{eqn:cvar_definition}
	\cvarvalue = \frac{1}{1-\alpha}\sum_{\{V \in \mathcal{V} | V \le \varvalue\}}\mathbb{P}[V]\cdot V,
\end{equation}
with $\varvalue$ given by Eq.~\eqref{eqn:var_definition}.
The \cvar{} of the portfolio at the $\alpha$ confidence level is then given by $\textrm{CVaR}_{\alpha}[V] \equiv V_0 - \cvarvalue$.

Classically, the $\textrm{VaR}_{\alpha}$ of a portfolio can be estimated by pricing the portfolio under each of the $N$ scenarios, sorting the resulting prices from smallest to largest, and picking the value at the $1-\alpha$ percentile as the \var{} estimate.
When the value of the derivative portfolio is computed classically using Monte Carlo to accuracy $\epsilon_p$, the \var{} computation requires separate Monte Carlo pricings for each scenario, and therefore the complexity of the algorithm scales as $\bigO{N/\epsilon_p^2}$.

A quantum algorithm for computing the \var{} of financial assets was first presented in Ref.~\cite{Woerner_2019} and later refined and applied to the use case of estimating capital requirements in order to manage counterparty credit risk in Ref.~\cite{egger2019credit}.
The algorithm seeks to estimate the \var{} of a financial asset at the $\alpha$ confidence level, through direct access to a unitary modeling the potential loss as a random variable $L$.
This unitary $\mathcal{L}$ creates a superposition

\begin{equation}
	\label{eqn:loss_unitary}
	\mathcal{L} : \ket{\vec{0}} \rightarrow \sum_{\ell}\sqrt{p_{\ell}}\ket{\ell}
\end{equation}
where the $\ket{\ell}$ register spans over possible values of $L$ encoded in an appropriate binary representation, and $p_{\ell}$ the corresponding probability of occurrence.
Picking a \var{} candidate $l$ and applying a unitary $\mathcal{C}$ which performs the binary comparison

\begin{equation}
	\label{eqn:comparator}
	\mathcal{C} : \ket{\ell}\ket{0} \rightarrow \ket{\ell}\ket{\ell < l},
\end{equation}
we get the state

\begin{equation}
\sum_{\ell \le l}\sqrt{p_{\ell}}\ket{\ell}\ket{0} + \sum_{\ell>l}\sqrt{p_{\ell}}\ket{\ell}\ket{1}.
\end{equation}
The probability of measuring the last qubit in the $\ket{0}$ state gives us the probability that the loss $L$ is smaller than $l$, $\mathbb{P}[\ket{0}]=\mathbb{P}[L \le l]$.
Using QAE to estimate $\mathbb{P}[\ket{0}]$ and applying a bisection search to find the smallest $l_{\alpha}$ such that $\mathbb{P}[L \le l_{\alpha}] \ge 1-\alpha$ gives us the \var{} of the asset at the $\alpha$ confidence level, $l_{\alpha}=\textrm{VaR}_{\alpha}[L]$.

This quantum method is not directly applicable to the calculation of derivative \var{} through the evaluation of scenario prices because in that case we do not have direct access to the unitary of Eq.~\eqref{eqn:loss_unitary}.
In the following section we show how to extended it in order to also be able to compute the \var{} of derivatives by incorporating one additional application of QAE.

\section{Value At Risk with Quantum Amplitude Estimation}
\label{sec:qae_var}
Quantum methods for the \var{} calculation of financial derivatives rely on the existence of an oracle $\mathcal{A}$ that encodes the price of a derivative $V$ into an amplitude

\begin{equation}
	\label{eqn:A_operator}
    \mathcal{A} : \ket{\vec{0}} \rightarrow \left( \sqrt{V}\ket{\psi_0}\ket{0} + \sqrt{1-V}\ket{\psi_1}\ket{1} \right),
\end{equation}
for arbitrary, normalized states $\ket{\psi_0}$, $\ket{\psi_1}$.
Different ways of constructing this oracle have been proposed in the literature \cite{rebentrost2018quantum, Stamatopoulos_2020, chakrabarti2021threshold}.
In Appendix~\ref{app:ae}, we describe in more detail the method proposed in \cite{chakrabarti2021threshold} and show how it can be extended to handle the case where we want Eq.~\eqref{eqn:A_operator} to encode the value of a portfolio of derivatives instead of the price of a single derivative.

\begin{figure}[t]
  \centering
  \includegraphics[width=0.7\linewidth]{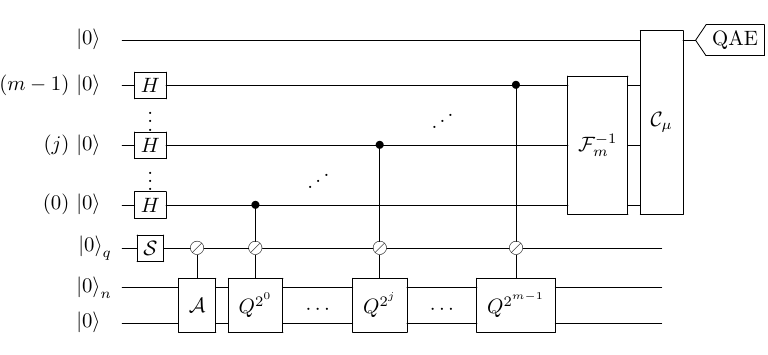}
  \caption{Quantum circuit required to prepare the state in Eq.~\eqref{eqn:qae_final_state}. Given a \var{} candidate $\mu$, the probability that the portfolio value is smaller than $\mu$ is encoded into the amplitude of a qubit being in the $\ket{0}$ state. First, the $\mathcal{S}$ and $\mathcal{A}$ operators of Eq.~\eqref{eqn:S_operator} and Eq.~\eqref{eqn:A_operator_scenario} respectively generate the superposition of all scenarios and encode the value of the portfolio under each scenario into the amplitude of the bottom qubit being in the $\ket{0}$ state. The $\oslash$ symbol indicates that an entire register acts as control to a unitary. Then, repeated controlled applications of an operator $\mathcal{Q}$ are applied with $\mathcal{Q}=\mathcal{A} S_{0}\mathcal{A}^{\dagger}S_{1}$, $S_{0}=\identity - 2\ket{\vec{0}}\bra{\vec{0}}$, $S_1=\identity-2\ket{1}\bra{1}$ where $S_0$ reflects about the bottom two registers being all in the $\ket{0}$ state and $S_1$ reflects around the bottom qubit in the $\ket{1}$ state. After an inverse Quantum Fourier Transform, the $m$ qubits acting as controls to the $\mathcal{Q}$ applications contain an $m$-qubit approximation to the portfolio value under each scenario in superposition, as shown in Eq.~\eqref{eqn:inner_qae}. Finally, a comparator circuit $C_{\mu}$ compares the binary value $V$ encoded in the $m$ qubits to $\mu$ and flips the top qubit if $V>\mu$, generating Eq.~\eqref{eqn:qae_final_state}. Estimating the probability of the top qubit being in the $\ket{0}$ state using QAE and re-running this circuit for different values of $\mu$ looking for the smallest value which gives $\mathbb{P}[\ket{0}]\ge 1-\alpha$ will give us a \var{} estimate at the $\alpha$ confidence level.}
  \label{fig:qae_var_circuit}
\end{figure}

In order to compute the \var{} of financial derivatives in the way described in Sec.~\ref{sec:derivative_var}, we construct scenarios consisting of tweaks to (classical) market data parameters which can be encoded in some appropriate way as a computational basis state $\ket{s}$.
For example, a scenario which simultaneously tweaks $d$ underlying parameters (also called \emph{risk factors}), can be encoded as

\begin{equation}
	\ket{s}=\ket{t_1 t_2, \dots t_d},
\label{eqn:scenario_encoding}
\end{equation}
where $t_i$ is a binary encoding of the tweak amount to underlying $i$.
For such scenario encoding $\ket{s}$, we can generalize the form of operator $\mathcal{A}$ of Eq.~\eqref{eqn:A_operator}, such that it encodes the value of the portfolio under scenario $s$

\begin{equation}
	\label{eqn:A_operator_scenario}
    \mathcal{A} : \ket{s}_q\ket{0}_n\ket{0} \rightarrow \ket{s}_q\left(\sqrt{V(s)}\ket{\psi_0^s}_n\ket{0} + \sqrt{1-V(s)}\ket{\psi_1^s}_n\ket{1} \right),
\end{equation}
for normalized quantum states $\ket{\psi_0^s}_n$ and $\ket{\psi_1^s}_n$.
Let $\mathcal{S}$ be an operator which loads $N$ scenarios $[ s_1, s_2, \dots, s_N]$ encoded this way in superposition, weighted by a probability denoting the likelihood of its occurrence

\begin{equation}
	\label{eqn:S_operator}
	\mathcal{S}: \ket{0}_q \rightarrow \sum_{i=0}^{N-1}\sqrt{p(s_i)}\ket{s_i}_q.
\end{equation}
The operator $\AS \equiv \mathcal{A}\mathcal{S}$ then encodes the portfolio value under all scenarios in superposition

\begin{equation}
	\label{eqn:AS_operator}
	\AS \equiv \mathcal{A}\mathcal{S}: \ket{\vec{0}} \rightarrow \sum_{i=0}^{N-1}\sqrt{p(s_i)}\ket{s_i}_q \left(\sqrt{V(s_i)}\ket{\psi_0^{s_i}}_n\ket{0} + \sqrt{1-V(s_i)}\ket{\psi_1^{s_i}}_n\ket{1} \right).
\end{equation}
Let us for now assume that each value $V(s_i)$ can be exactly represented using $m$ qubits.
The circuitry of QAE (up to measurement) can then be applied with the marked state identified as the last qubit being in the $\ket{0}$ state to get a binary representation of each $V(s_i)$ into a quantum register

\begin{equation}
	\label{eqn:inner_qae}
	\sum_{i=0}^{N-1}\sqrt{p(s_i)}\ket{s_i}_q\ket{V(s_i)}_m \ket{\textrm{garbage}}
\end{equation}
where the state of the remaining registers $\ket{\textrm{garbage}}$ will depend on the choice of QAE variant used, but is not involved in the remainder of the process and can be ignored.
Then the comparator of Eq.~\eqref{eqn:comparator} can be applied to the $\ket{V(s_i)}$ register with a \var{} candidate $\mu$ and an ancilla qubit initially in the $\ket{0}$ state to obtain

\begin{equation}
	\label{eqn:qae_final_state}
	\sum_{\{i  \: | \: V(s_i) \le \mu\}}\sqrt{p(s_i)}\ket{s_i}_q\ket{V(s_i)}_m\ket{0} + \sum_{\{i  \: | \: V(s_i) > \mu\}}\sqrt{p(s_i)}\ket{s_i}_q\ket{V(s_i)}_m\ket{1}.
\end{equation}
The probability of the last qubit being in the $\ket{0}$ state gives us the probability that the portfolio value $V$ is smaller that $\mu$ across all scenarios considered, $\mathbb{P}[\ket{0}]=\mathbb{P}[V \le \mu]$.
We can thus employ consecutive rounds of QAE to find the smallest $\mu_{\alpha}$ such that $\mathbb{P}[V \le \mu_{\alpha}] \ge 1-\alpha$, which can be achieved with a bisection search over values of $\mu$.
Assuming the portfolio value today is $V_0$, the \var{} at the $\alpha$ confidence level will be given by $\textrm{VaR}_{\alpha}[V] \equiv V_0 - \mu_{\alpha}$.
Note that if we use the \emph{canonical} amplitude estimation \cite{brassard2002quantum} circuitry to generate Eq.~\eqref{eqn:inner_qae}, the value encoded in the resulting register will not be $V(s_i)$ but rather $\theta_i = \arcsin(\sqrt{V(s_i)})$ \footnote{Canonical QAE gives an equal superposition of two values $\theta_1 = \arcsin(\sqrt{V}) \in [0, \pi/2]$ and $\theta_2 = \pi - \theta_1 \in [\pi/2, \pi]$. Quantum arithmetic can be used to perform $\theta \rightarrow \pi - \theta$ if $\theta \in [\pi/2, \pi]$ which will ensure the register represents a value $\theta \in [0, \pi/2]$ such that $V = \sin^2(\theta)$.}.
While quantum arithmetic can be used to compute a binary approximation to $V(s_i)$ from $\theta_i$, since we are only interested performing the comparison $V(s_i) \le \mu$ with $V(s_i) \in [0,1]$, we can adjust the comparator circuit to equivalently perform $\theta_i \le \arcsin(\sqrt{\mu})$.
A diagram of the circuit required to generate Eq.~\eqref{eqn:qae_final_state} using canonical QAE is shown in Fig.~\ref{fig:qae_var_circuit}.

In practice, we will not be able to represent all values $V(s_i)$ exactly in binary, and instead of Eq.~\eqref{eqn:inner_qae}, for each scenario value we will get a superposition over all $M=2^m$ possible states representable using $m$ qubits, where each state will be weighed by a probability depending on the value of $V(s_i)$

\begin{equation}
	\label{eqn:inner_qae_superposition}
	\sum_{i=0}^{N-1}\sqrt{p(s_i)}\ket{s_i}_q\sum_{j=0}^{M-1}\sqrt{q_j(V(s_i))}\ket{\tilde{V}_j}_m,
\end{equation}
with the probabilities given by \cite{brassard2002quantum}

\begin{equation}
	\label{eqn:qae_probabilities}
	q_j(x) = \frac{\sin^2\left(M(\tilde{V}_j-x)\pi\right)}{M^2\sin^2\left((\tilde{V}_j-x)\pi\right)},
\end{equation}
which will peak around the values $\tilde{V}_j$ closest to each $V(s_i)$.
Applying the comparator unitary to perform $\ket{\tilde{V}_j}\ket{0}\rightarrow\ket{\tilde{V}_j}\ket{\tilde{V}_j \le \mu}$ will give

\begin{equation}
	\sum_{i=0}^{N-1}\sqrt{p(s_i)}\ket{s_i}_q \left(\sum_{\{j  \: | \: \tilde{V}_j \le \mu\}}\sqrt{q_j(V(s_i))}\ket{\tilde{V}_j}_m\ket{0} + \sum_{\{j  \: | \: \tilde{V}_j > \mu\}}\sqrt{q_j(V(s_i))}\ket{\tilde{V}_j}_m\ket{1}\right),
\end{equation}
meaning that every scenario value $V(s_i)$ will (incorrectly) contribute something to the probability of $\ket{1}$ if $V(s_i) \le \mu$ and to $\ket{0}$ if $V(s_i) > \mu$.
Increasing the number of qubits used for this \emph{inner} application of QAE will limit the probability that is erroneously ``allocated" to the wrong flag state.

Note that the QAE method used to produce Eq.~\eqref{eqn:inner_qae_superposition} needs to coherently approximate $V(s_i)$, meaning highly performant iterative QAE methods \cite{grinko2021iterative, suzuki2020amplitude, Giurgica_Tiron_2022} cannot be employed for this task.
However, the original QAE formulation \cite{brassard2002quantum} as well as the QAE variant introduced in Ref.~\cite{Rall_2021} can be used for coherent estimation.
While the former requires the Quantum Fourier Transform (QFT) and overhead in terms of ancilla qubits, the latter can produce the approximation coherently with comparable performance to the iterative versions.

\section{Value At Risk with Quantum Signal Processing}
\label{sec:qsp_var}
\subsection{Quantum Signal Processing}
Quantum Signal Processing is a technique which performs polynomial transformations to the singular values of a matrix $A$ that has been \emph{block-encoded} into a unitary operator \footnote{This technique was originally introduced as Quantum Singular Value Transformation in Ref.~\cite{gilyen2019quantum}, but here we use the nomenclature of Ref.~\cite{martyn2021grand}, which refers to the underlying method as Quantum Signal Processing, encompassing a variety of applications.}.
Specifically, given projectors $\Pi$ and $\tilde{\Pi}$, we say that a unitary $U$ acting on $n$ qubits is a block-encoding of a matrix $A$ if $A=\tilde{\Pi}U\Pi$, such that the projectors $\Pi$ and $\tilde{\Pi}$ determine the location of $A$ in $U$

\begin{eqnarray}
	\label{eqn:block_encoding}
	U = \kbordermatrix{\mbox{} &\Pi &  \\
	\tilde{\Pi} & A     & \cdot \\
	& \cdot & \cdot
	}.
\end{eqnarray}
QSP consists of consecutive applications of $U$ and $U^\dagger$, interleaved with projector-controlled rotation operators $\Pi_{\phi}=e^{i\phi(2\Pi - \identity)}$ and $\tilde{\Pi}_{\phi}=e^{i\phi(2\tilde{\Pi} - \identity)}$, which for phase factors $\vec{\phi}=(\phi_1, \phi_2, \cdots, \phi_d)$ induce a polynomial transformation on $A$ such that \cite{martyn2021grand, gilyen2019quantum}

\begin{equation}
	A=\sum_{k}\sigma_k\ket{w_k}\bra{v_k} \rightarrow P(A)=\sum_{k}P(\sigma_k)\ket{w_k}\bra{v_k},
\end{equation}
where $\sigma_k$ denote the singular values of $A$ while $\{\ket{w_k}\}$, $\{\ket{v_k}\}$ are the left and right singular vectors of $A$ respectively, and $P$ is a $d$-degree polynomial.
When $A$ is Hermitian, in which case the singular values of the matrix coincide with its eigenvalues, Quantum Signal Processing effectively performs a polynomial transformation to the eigenvalues of the matrix.

More formally, define the operators

\begin{align}
	\label{eqn:qsp_basic_unitary}
	\mathcal{U}^{\vec{\phi}} = \begin{cases}
					\tilde{\Pi}_{\phi_1}U\displaystyle \prod_{k=1}^{(d-1)/2}\Pi_{\phi_{2k}}U^{\dagger}\tilde{\Pi}_{\phi_{2k+1}}U, & \text{for odd $d$,} \\
					\displaystyle\prod_{k=1}^{d/2}\Pi_{\phi_{2k-1}}U^{\dagger}\tilde{\Pi}_{\phi_{2k}}U, & \text{for even $d$}
	\end{cases}
\end{align}

\begin{equation}
	\label{eqn:qsp_conditional_unitary}
	\mathcal{U}_C^{\vec{\phi}} = \left(\mathcal{U}^{\vec{\phi}} \otimes \ket{0}\bra{0} +  \mathcal{U}^{-\vec{\phi}} \otimes \ket{1}\bra{1} \right).
\end{equation}
If $U$ is an $n$-qubit unitary, the polynomial transformation achieved through QSP is implemented with \cite{martyn2021grand}

\begin{equation}
	\label{eqn:QSP_general_transformation}
U^{\vec{\phi}} = (\identity^{\otimes n} \otimes H)  \mathcal{U}_C^{\vec{\phi}}  (\identity^{\otimes n} \otimes H) =
	\kbordermatrix{\mbox{} & \big(\Pi \otimes \ket{0}\bra{0} \big) &  \\
	\left(\Pi' \otimes \ket{0}\bra{0} \right) & P(A)     & \cdot \\
	& \cdot & \cdot
	}
\end{equation}
where $\Pi'=\tilde{\Pi}$ for odd $d$ and $\Pi'=\Pi$ for even $d$.

In practice, $d$ invocations of $U$ perform a $d$-degree polynomial transformation to $A$ and the choice of phase factors $\vec{\phi}=(\phi_1, \phi_2, \cdots, \phi_d)$ determine the exact polynomial that is applied.
Remarkably, the reverse is also true:
for $a \in [-1, 1]$ and any real polynomial $P \in \mathbb{R}(a)$, there exists a sequence of QSP phase factors $\vec{\phi}=(\phi_1, \phi_2, \cdots, \phi_d)$ for which Eq.~\eqref{eqn:QSP_general_transformation}  holds, as long as $\textrm{deg}(P) \le d$, $|P(a)| \le 1, \forall a \in [-1, 1]$ and $P$ either even or odd.
Additionally, for any function $f(a)$ satisfying these conditions we can first identify a polynomial approximation to the function, determine the QSP phase factors $\vec{\phi}$ for that polynomial, and use QSP to approximately apply $f(a)$ to the singular values of the block-encoded matrix.
Various methods have been proposed to generate polynomial approximations to generic functions, where the approximation is either constructed analytically \cite{martyn2021grand, Haah2019product, chao2020finding} or with optimization-based numerical methods \cite{dong2021efficient, dong2022ground}.

\subsection{Algorithm for Value at Risk}
\label{sec:qsp_algorithm}
Defining $\tilde{\Pi} \equiv I^{\otimes (n+q)} \otimes \ket{0}\bra{0} $ and $\Pi \equiv I^{\otimes q} \otimes \left(\ket{0}\bra{0} \right)^{\otimes (n+1)}$, observe that $\tilde{\Pi}\mathcal{A}\Pi$ with $\mathcal{A}$ given by Eq.~\eqref{eqn:A_operator_scenario}, is a rank-1 matrix with a single non-trivial singular value $\sqrt{V(s)}$ \cite{gilyen2019quantum}.
Therefore, we can use QSP with the operator $\mathcal{A}$ acting as the block encoding unitary of Eq.~\eqref{eqn:block_encoding} to apply polynomial transformations to the values $\sqrt{V(s)}$ as discussed in the previous section.
After the superposition of scenarios is created with the unitary $\mathcal{S}$ of Eq.~\eqref{eqn:S_operator}, the unitary $U^{\vec{\phi}}$ of Eq.~\eqref{eqn:QSP_general_transformation} based on $\mathcal{U}^{\vec{\phi}}$ with $U=\mathcal{A}$ produces the state
\begin{equation}
	\label{eqn:qsp_derivative_transformation}
	U^{\vec{\phi}} \mathcal{S} \ket{0}_q \ket{0}_{n+1}\ket{0} = \sum_{i=0}^{N-1}\sqrt{p(s_i)}\ket{s_i}_q
	\begin{cases}
			\left(P(\sqrt{V(s_i)})\ket{\psi_0}_n\ket{0}_2 + \sqrt{1-P(\sqrt{V(s_i)})^2}\ket{\psi_1}_{n}\ket{0_{\perp}}_2 \right), & \text{for odd $d$,} \\
			& \\
			\left(P(\sqrt{V(s_i)})\ket{0}_{n+2} + \sqrt{1-P(\sqrt{V(s_i)})^2}\ket{0_{\perp}}_{n+2} \right), & \text{for even $d$,}
	\end{cases}
\end{equation}
for normalized quantum states $\ket{\psi_0}_n$, $\ket{\psi_1}_n$, and $\ket{0_{\perp}}_{a}$ denoting a normalized state orthogonal to $\ket{0}_{a}$.
The polynomial $P(x)$ is determined by the choice of phase factors $\vec{\phi}$.

If we can find phase factors such that $P(x)$ is an approximation to the threshold function

\begin{equation}
	\label{eqn:threshold_function}
	\theta_{\mu}(x) = \begin{cases}
1, \; {x} \le {\mu} \\
0, \; {x} > {\mu},
\end{cases}
\end{equation}
Eq.~\eqref{eqn:qsp_derivative_transformation} will give us

\begin{equation}
	\label{eqn:qsp_var_final_state}
	\sum_{\{i  \: | \: \sqrt{V(s_i)} \le \mu\}}\sqrt{p(s_i)}\ket{G} + \sum_{\{i  \: | \: \sqrt{V(s_i)} > \mu\}}\sqrt{p(s_i)}\ket{G_{\perp}},
\end{equation}
where $\ket{G}=\ket{0}_2$ for odd $d$ and $\ket{G}=\ket{0}_{n+2}$ for even $d$, ignoring the (normalized) state of the remaining registers for clarity.
The probability of measuring $\ket{G}$ will then be equal to the probability that the portfolio value across all scenarios is smaller than $\mu^2$

\begin{equation}
	\label{eqn:threshold_prob}
	\mathbb{P}(\ket{G}) = \sum_{ \{s  \: | \: V(s) \le \mu^2\} } p(s).
\end{equation}
Therefore, the smallest value of $\mu = \mu_{\alpha}$ for which the probability in Eq.~\eqref{eqn:threshold_prob} satisfies $\mathbb{P}(\ket{G}) \ge 1-\alpha$ gives us the value of $\varvalue=\mu_{\alpha}^2$ we are looking for in Eq.~\eqref{eqn:var_definition} in order to estimate the VaR of the portfolio at the $\alpha$ confidence level.
We can find such $\mu_{\alpha}$ by performing bisection search over values of $\mu \in [0, 1]$ and applying QAE to estimate the probability in Eq.~\eqref{eqn:threshold_prob} in each iteration.
The bisection search ends when we find a value of $\mu$ for which our estimate of $\mathbb{P}(\ket{G})$ is $\epsilon$-close to $1-\alpha$ for some value of $\epsilon$ determined by the target accuracy of the QAE method employed.
By the union bound, if each round of QAE is performed at confidence level $1-\alpha_{A}$, the confidence of our final \var{} estimate after $k$ rounds will be $1-k\cdot\alpha_{A}$.

\begin{figure}[h]
  \centering
  \includegraphics[width=0.5\linewidth]{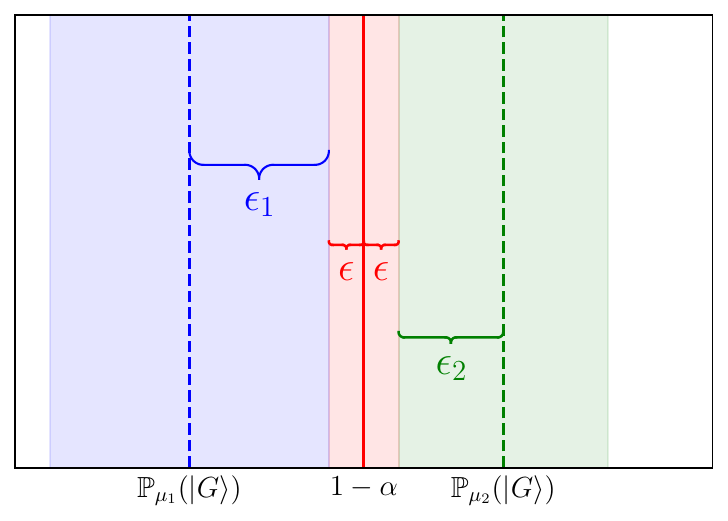}
  \caption{For each round of  bisection search in Algorithm~\ref{algo:qsp_var}, we are trying to determine whether for a given value $\mu$, the value $\mathbb{P}(\ket{G})$ of Eq.~\eqref{eqn:threshold_prob} is $\epsilon$-close to $1-\alpha$. However, the accuracy with which we estimate the value $\mathbb{P}(\ket{G})$ can be relaxed depending on how far $\mathbb{P}(\ket{G})$ is from $1-\alpha$. For example, for the values $\mathbb{P}_{\mu_1}(\ket{G})$ and $\mathbb{P}_{\mu_2}(\ket{G})$ corresponding to some thresholds $\mu_1, \mu_2$ shown above, each probability only needs to be estimated with accuracy $\epsilon_1, \epsilon_2 > \epsilon$ respectively in order to proceed with the bisection search.}
  \label{fig:adaptive_var_epsilon}
\end{figure}

This approach can be optimized by observing that not every bisection step needs to use the same accuracy $\epsilon$ or confidence level parameter $\alpha_{A}$ for QAE.
Because at every step we need to estimate whether the value $\mathbb{P}(\ket{G})$ in Eq.~\eqref{eqn:threshold_prob} is $\epsilon$-close to $1-\alpha$, we can employ QAE with accuracy $\epsilon_k > \epsilon$ in cases where $\mathbb{P}(\ket{G})$ is significantly far from $1-\alpha$.
Using iterative amplitude estimation routines such as IQAE \cite{grinko2021iterative}, we can adaptively increase the accuracy of the estimate by taking progressively more samples until we can determine whether $|\mathbb{P}(\ket{G}) - (1 -\alpha)| \le \epsilon$ with some confidence $1-\alpha_{A}$ and then decide on the next bisection search step accordingly.
This observation is illustrated in Fig.~\ref{fig:adaptive_var_epsilon}.

\begin{figure}[t]
  \centering
  \includegraphics[width=0.85\linewidth]{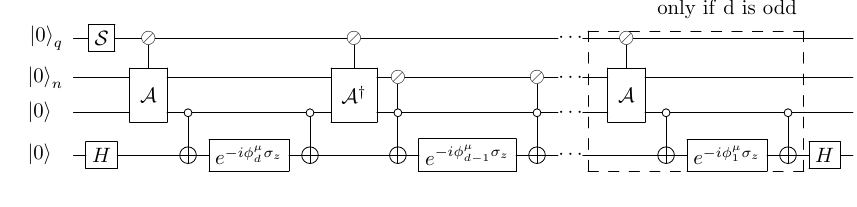}
  \caption{Circuit diagram depicting one iteration of the QSP-based \var{} estimation method described algorithm in Sec.~\ref{sec:qsp_algorithm} and outlined in Algorithm~\ref{algo:qsp_var}. Given a \var{} candidate $\mu$, the probability that the portfolio value is smaller than $\mu^2$ is encoded into the amplitude of a marked state. The $\mathcal{S}$ operator of Eq.~\eqref{eqn:S_operator} is first applied to generate the superposition of all scenarios considered.
  Then, a $d$-degree polynomial approximation to the threshold function $\theta_{\mu}(x)$ of Eq.~\eqref{eqn:threshold_function} along with the corresponding phase factors $\vec{\phi}^{\mu} = (\phi_1^{\mu}, \phi_2^{\mu}, \cdots, \phi_d^{\mu})$ are computed, and the generated polynomial transformation is implemented using the QSP unitary of Eq.~\eqref{eqn:qsp_derivative_transformation}, which consists of alternating applications of controlled phase rotations and the block-encoding operator $\mathcal{A}$ of Eq.~\eqref{eqn:A_operator_scenario}.
  After the transformation, the probability of measuring $\ket{G}=\ket{0}_2$ in the bottom two qubits if $d$ is odd or $\ket{G}=\ket{0}_{n+2}$ in the bottom $n+2$ qubits if $d$ is even, gives us the probability that the portfolio value across all scenarios is smaller than $\mu^2$ as shown in Eq.~\eqref{eqn:threshold_prob}.
  Estimating this probability using QAE and re-running this circuit for different values of $\mu$ looking for the smallest value which gives $\mathbb{P}[\ket{G}]\ge 1-\alpha$ will give us a \var{} estimate at the $\alpha$ confidence level.}
  \label{fig:qsp_var_circuit}
\end{figure}

The confidence level $1-\alpha_{A}$ of QAE can similarly be adjusted in each round.
If the target confidence level of the \var{} estimation is denoted as $\textrm{CL}$, by the union bound the confidence level $\alpha_k$ in the $k$-th QAE invocation can be adjusted as long as $\textrm{CL} = 1- \sum_{k}\alpha_k$ is satisfied.
Therefore, the total probability of failure $1-\textrm{CL}$ can be distributed across each QAE round as desired.
One possible choice is to allocate higher probability of failure when the QAE target error $\epsilon_k$ is small, which would decrease the constant factor of the $\bigO{1/{\epsilon_k}}$ complexity of QAE when $1/{\epsilon_k}$ is large.
The pseudocode for this QSP-based algorithm is given in Algorithm~\ref{algo:qsp_var} and the circuit required to generate Eq.~\eqref{eqn:threshold_prob} is shown in Fig.~\ref{fig:qsp_var_circuit}.

\SetKwInput{Input}{Input}
\SetKwInput{Output}{Output}
\SetKw{KwTo}{in}\SetKwFor{For}{for}{\string:}{}%
\SetKwIF{If}{ElseIf}{Else}{if}{:}{elif}{else:}{}%
\SetKwFor{While}{while}{:}{fintq}%

\AlgoDontDisplayBlockMarkers\SetAlgoNoEnd\SetAlgoNoLine%

\begin{algorithm} 
	\label{algo:qsp_var}
\caption{VaR Estimation using Quantum Signal Processing}
\DontPrintSemicolon
\Input{Amplitude Estimation accuracy $\epsilon_{A}$}
\Input{VaR confidence level $\alpha$}
Set $\mu_l=0$ and $\mu_h=1$\;
Set $k=0$\;
\While{True}
{
	Apply the $S$ operator of Eq.~\eqref{eqn:S_operator} to generate the superposition of all scenarios used to compute the \var{} of a derivative portfolio whose value today is $V_0$
	\begin{equation*}
		\sum_{i=0}^{N-1}\sqrt{p(s_i)}\ket{s_i}
	\end{equation*}\;
	Pick $\mu = \mu_l + (\mu_h - \mu_l)/2$ \;
	Generate a $d$-degree polynomial and corresponding phase factors $\vec{\phi}^{\mu} = (\phi_1^{\mu}, \phi_2^{\mu}, \cdots, \phi_d^{\mu})$ such that the polynomial transformation $P(A)$ in Eq.~\eqref{eqn:QSP_general_transformation} is an approximation to the threshold function $\theta_{\mu}(x)$ of Eq.~\eqref{eqn:threshold_function}. \;
	Apply the generated polynomial transformation using QSP and the block-encoding operator $\mathcal{A}$ of Eq.~\eqref{eqn:A_operator_scenario} such that the probability of the last qubits being in the $\ket{G}$ state with $\ket{G}=\ket{0}_2$ for odd $d$ and $\ket{G}=\ket{0}_{n+2}$ for even $d$ becomes
	\begin{equation*}
		\mathbb{P}(\ket{G}) = \sum_{ \{s  \: | \: V(s) \le \mu^2\} } p(s)
	\end{equation*}\;
	Iteratively estimate the probability $\mathbb{P}(\ket{G})$ using QAE with decreasing estimation error $\epsilon_k \ge \epsilon_A$  to determine bounds satisfying $\mathbb{P}(\ket{G}) \in [p_l, p_h]$ with $p_h-p_l \le 2\epsilon_k$ at confidence level $1-\alpha_k$, until we either determine that $1 - \alpha \notin [p_l, p_h]$ or the target estimation error reaches $\epsilon_k = \epsilon_A$. \;
	Set $k = k + 1$ \;
	\If{$p_h < 1-\alpha$}
		{Set $\mu_l = \mu$}
	\ElseIf{$p_l > 1-\alpha$}
		{Set $\mu_h = \mu$}
	\Else
		{Return $\textrm{VaR}_{\alpha}$ estimate $V_0-\mu^2$ with confidence $1- \sum_{k}\alpha_k$}
}
\end{algorithm}

\subsection{Conditional Value at Risk}
Once we have estimated the \var{}, the \cvar{} of derivatives can be also be estimated using QSP by applying a modified threshold function which is linear below the threshold and zero above

\begin{equation}
	\label{eqn:linear_threshold_function}
	\tilde{\theta}_{\mu}(x) = \begin{cases}
x, \; {x} \le {\mu} \\
0, \; {x} > {\mu},
\end{cases}
\end{equation}
at $\mu = \mu_{\alpha}$ where $\mu_{\alpha}$ is the threshold estimated in the \var{} calculation such that the probability in Eq.~\eqref{eqn:threshold_prob} equals $1-\alpha$.
Using QSP to apply the transformation of Eq.~\eqref{eqn:qsp_derivative_transformation} with $P(x)$ now a polynomial approximation to this function gives us

\begin{equation}
	\label{eqn:qsp_cvar_final_state}
	\sum_{\{i  \: | \: \sqrt{V(s_i)} \le \mu_{\alpha}\}}\sqrt{p(s_i)}\sqrt{V(s_i)}\ket{G} + \sum_{\{i  \: | \: \sqrt{V(s_i)} \le \mu_{\alpha}\}}\sqrt{p(s_i)}\sqrt{1-V(s_i)}\ket{G_{\perp}} + \sum_{\{i  \: | \: \sqrt{V(s_i)} > \mu\}}\sqrt{p(s_i)}\ket{G_{\perp}}.
\end{equation}
Applying QAE to estimate the probability of measuring $\ket{G}$ will give us the expected value of the portfolio below the \var{} level

\begin{equation}
	\label{eqn:threshold_prob_cvar}
	\mathbb{P}(\ket{G}) = \sum_{ \{s  \: | \: V(s) \le \mu_{\alpha}^2\} } p(s)V(s).
\end{equation}
Dividing this quantity by $\mathbb{P}[V \le \mu_{\alpha}^2]$ will give us $\cvarvalue$ of Eq.~\eqref{eqn:cvar_definition} and the \cvar{} of the portfolio is then given by $V_0-\cvarvalue$, if the portfolio value today is $V_0$.

Because the threshold function of Eq.~\eqref{eqn:linear_threshold_function} is discontinuous at $x=\mu$, we will need a high degree polynomial to generate a close approximation to the function.
We can however avoid the discontinuity at $x=\mu$ by inverting the threshold function in Eq.~\eqref{eqn:linear_threshold_function} in the region below $\mu$ such that

\begin{equation}
	\label{eqn:inverted_linear_threshold_function}
	\hat{\theta}_{\mu}(x) = \begin{cases}
\mu-x, \; {x} \le {\mu} \\
0, \; {x} > {\mu}.
\end{cases}
\end{equation}
Applying this transformation instead at $\mu = \mu_{\alpha}$ gives us

\begin{equation}
	\label{eqn:cvar_threshold_prob}
	\mathbb{P}(\ket{G}) = \sum_{ \{s  \: | \: V(s) \le \mu_{\alpha}^2\} } p(s)\left(\mu_{\alpha}-V(s)\right),
\end{equation}
from which we compute

\begin{equation}
	\hat{C}_{\alpha} = \frac{1}{\mathbb{P}[V \le \mu_{\alpha}^2]} \sum_{ \{s  \: | \: V(s) \le \mu_{\alpha}^2\} } p(s)\left(\mu_{\alpha}-V(s)\right) = \left(\frac{1}{\mathbb{P}[V \le \mu_{\alpha}^2]}\sum_{ \{s  \: | \: V(s) \le \mu_{\alpha}^2\} } p(s)\mu_{\alpha}\right) - \cvarvalue= \mu_{\alpha} -\cvarvalue.
\end{equation}
The \cvar{} can then be calculated as $\textrm{CVaR}_{\alpha}[V] = V_0 - \cvarvalue = V_0 - (\mu_{\alpha} - \hat{C}_{\alpha})$.

\subsection{Threshold Function Approximation using QSP}
\label{sec:thres_approximation}
In order to apply the threshold functions in Eq.~\eqref{eqn:threshold_function} and Eq.~\eqref{eqn:linear_threshold_function} using QSP, we first need to find polynomials which approximate them.
The polynomial transformations possible through the QSP framework must have definite parity (even or odd), and because we aim to transform real positive amplitudes representing derivative prices, we restrict our attention to the interval $[0,1]$ due to symmetry.
For the threshold function in in Eq.~\eqref{eqn:threshold_function} used for the \var{} calculation, we follow the method in Ref.~\cite{dong2022ground} and look for a real polynomial $f(x)$ satisfying

\begin{equation}
	|f(x) - c| \le \epsilon, \quad \forall x \in [0, \mu-\Delta/2]; \quad |f(x)| \le \epsilon, \quad \forall x \in [\mu+\Delta/2, 1]
\end{equation}
where $c$ is chosen close to $1$ but preferably slightly smaller to avoid overshooting, $\epsilon$ controls how far away the function is allowed to deviate from the values $0$ and $1$, and $\Delta$ is a gap parameter which controls the steepness of the jump from $1$ to $0$ at $\mu$ where $\theta_{\mu}(x)$ is discontinuous.
Smaller values of $\Delta$ will create a better approximation to the function in the interval $[\mu-\Delta/2, \mu+\Delta/2]$ and a smaller value of $\epsilon$ will correspondingly allow for a better approximation in the interval $[0, \mu-\Delta/2] \cup [\mu+\Delta/2, 1]$.

While there are analytical methods of obtaining an approximation $f(x)$ to this function with a polynomial of degree $\bigO{(1/\Delta)\log(1/\epsilon)}$ \cite{low2017hamiltonian}, we employ the optimization-based method described in Ref.~\cite{dong2022ground} which generates near-optimal approximations without relying on any analytic computation.
This method constructs a linear combination of even Chebyshev polynomials with some unknown coefficients $\{c_k\}$

\begin{equation}
	f(x)=\sum_{k=0}^{d/2}c_kT_{2k}(x),
\end{equation}
and aims to find the coefficients $\{c_k\}$ which give the best approximation to $\theta_{\mu}(x)$.
This process is formulated as a discrete optimization problem by discretizing the interval $[0, 1]$ using $M$ grid points generated by the roots of Chebyshev polynomials $\left\{x_j = -\cos\frac{j\pi}{M-1}\right\}_{j=0}^{M-1}$ and solving the following minimax optimization problem for the coefficients $\{c_k\}$

\begin{eqnarray}
	\label{eqn:threshold_optimization}
	&\underset{\{c_k\}}{\min} \quad \max \left\{ \underset{x_j \in [0, \mu-\Delta/2]}{\max} |f(x_j)-c|, \underset{x_j \in [\mu + \Delta/2, 1]}{\max} |f(x_j)| \right\} \nonumber \\
	& \textrm{s.t.} \quad f(x_j)=\sum_k c_kA_{jk}, \quad |f(x_j)| \le c, \quad \textrm{for } j=0, 1, \ldots, M-1,
\end{eqnarray}
where $A_{jk}$ is a matrix of coefficients defined as $A_{jk}=T_{2k}(x_j)$ for $k=0, 1, \ldots, d/2$.

\begin{figure}[t]
\subfloat[]{\label{fig:var_thres_approximation}\includegraphics[width=.49\linewidth]{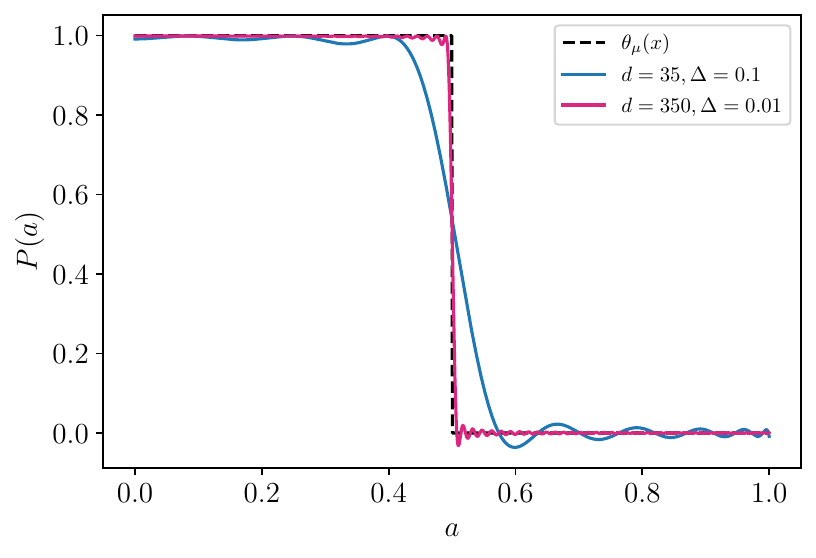} }%
\subfloat[]{\label{fig:cvar_thres_approximation}\includegraphics[width=.49\linewidth]{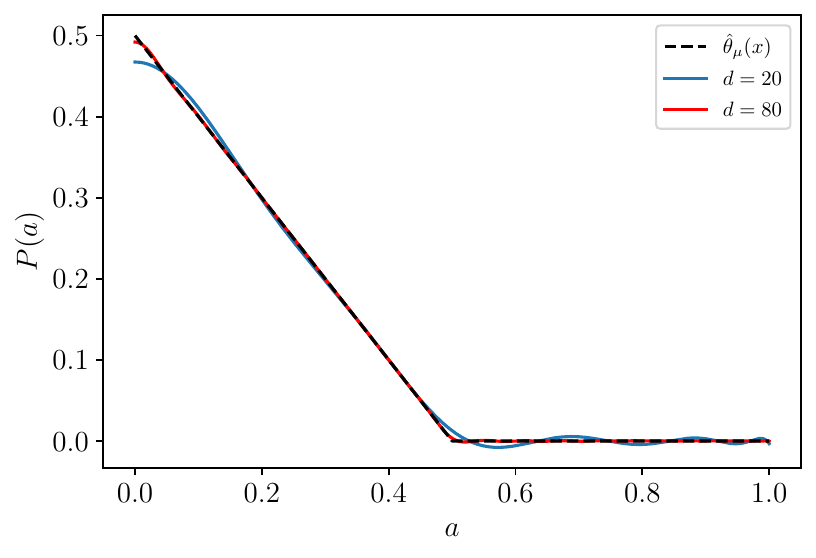} }%
\caption{Polynomial approximations generated using the optimization-based method described in Sec.~\ref{sec:thres_approximation}. Using Quantum Signal Processing we can perform the transformation $a \rightarrow P(a)$ for each singular value $a$ of a block-encoded unitary, where increasing the degree of the polynomial $P(a)$ will result in a better approximation to the target function. (a) Approximations to the threshold function of Eq.~\eqref{eqn:threshold_function} at $\mu=0.5$ used for the estimation of \var{}, generated by optimizing Eq.~\eqref{eqn:threshold_optimization} for different polynomial degrees $d$ and gap parameter $\Delta$, with $\epsilon=1-c=10^{-3}$. (b) Approximations to the linear threshold function of Eq.~\eqref{eqn:inverted_linear_threshold_function} used for the estimation of \cvar{}, generated using Eq.~\eqref{eqn:linear_threshold_optimization} for different polynomial degrees $d$, $\Delta=10^{-3}$ and $\epsilon=1-c=10^{-3}$. Avoiding the discontinuity at $a=\mu$ by using the function of Eq.~\eqref{eqn:inverted_linear_threshold_function} instead of Eq.~\eqref{eqn:linear_threshold_function} in the calculation of \cvar{} results in significantly lower degree polynomials required to approximate the function with high degree of accuracy.}%
\label{fig:thres_approximation}
\end{figure}

We follow the same procedure to construct a polynomial approximation to the inverted threshold function $\hat{\theta}_{\mu}$ of Eq.~\eqref{eqn:inverted_linear_threshold_function} used in the estimation of \cvar{}.
In this case, we solve the minimax optimization problem

\begin{eqnarray}
	\label{eqn:linear_threshold_optimization}
	\underset{\{c_k\}}{\min} \quad \max \left\{ \underset{x_j \in [0, \mu-\Delta/2]}{\max} |f(x_j)-(\mu - x_j)|, \underset{x_j \in [\mu + \Delta/2, 1]}{\max} |f(x_j)| \right\},
\end{eqnarray}
subject to the same constraints as Eq.~\eqref{eqn:threshold_optimization}.
The term $|f(x_j)-(\mu - x_j)|$ now enforces that the polynomial approximation is close to $\mu-x$ in the interval $[0, \mu-\Delta/2]$.
Polynomial approximations generated using the optimization-based method described in this section for the threshold functions $\theta_{\mu}(x)$ and $\hat{\theta}_{\mu}(x)$ at $\mu=0.5$ are shown in Fig.~\ref{fig:thres_approximation}.

While the calculation of \cvar{} is just as (or more) important than \var{} as a business use case, in practice they are both formulated similarly in the QSP framework.
In fact, computing the \cvar{} using QSP requires first the computation of \var{} and the correct threshold to be identified.
As such, in the following sections we focus on the performance and estimation errors of the \var{} algorithm which is the core component in the estimation of both risk metrics.

\section{Complexity And Error Analysis}
\label{sec:error_analysis}
The \var{} calculation methods discussed so far rely on estimating the probability that a portfolio value $X$ will be less than some threshold $\mu$.
In the continuous setting, this can be written as

\begin{equation}
	\label{eqn:threshold_integral}
	P[X \le \mu] = \int_0^1 \theta_{\mu}(x)p(x)dx,
\end{equation}
where $p(x)$ denotes the probability of value $x$ occurring, $\theta_{\mu}(x)$ is the threshold function defined in Eq.~\eqref{eqn:threshold_function} and the portfolio values have been normalized to lie in $[0,1]$.
The estimation of \var{} is then the solution to the inverse problem, where given a confidence level $\alpha$, we try to determine the value $\mu_{\alpha}$ giving $P[X \le \mu_{\alpha}] \ge 1-\alpha$, which can be formulated as finding the root of

\begin{equation}
	\label{eqn:inverse_root}
	\tilde{f}(\mu_{\alpha}) = (1-\alpha) - P[X \le \mu_{\alpha}].
\end{equation}
The error in the estimation of \var{} will be determined by the accuracy with which we can approximate Eq.~\eqref{eqn:threshold_integral} and the accuracy in finding the root of Eq.~\eqref{eqn:inverse_root}, given the approximation to $P[X \le \mu_{\alpha}]$.
An error $\delta$ in the estimate of $P[X \le \mu_{\alpha}]$ will result in error $\epsilon$ in the estimate of the \var{} given by

\begin{equation}
	\epsilon=\delta \cdot \left. \frac{dF^{-1}(p)}{dp}\right\rvert_{p=1-\alpha} + \bigO{\delta^2},
\end{equation}
where $F^{-1}(p)$ is the inverse cumulative distribution function of the distribution $p(x)$, assuming $F^{-1}(p)$ is analytic at $p = 1 - \alpha$.
We now examine the accuracy with which $P[X \le \mu_{\alpha}]$ can be estimated by various methods, which to first order gives the accuracy of the \var{} estimate (with a constant factor depending on the shape of the probability distribution $p(x)$ at $\mu_{\alpha}$).

\subsection{Classical and Semi-Classical Methods}
Classical methods for the \var{} estimation of financial derivatives use two nested Monte Carlo invocations; an inner round to estimate the price of the derivative $x$ (see Eq.~\eqref{eqn:derivative_exp_value}), and an outer round sampling scenarios to estimate the integral of Eq.~\eqref{eqn:threshold_integral}.
When $N$ scenarios are used to approximate the integral and the derivative price for each scenario is evaluated to accuracy $\epsilon_p$, the approximation $P'$ can be written as

\begin{equation}
	P'[X \le \mu] = \frac{1}{N}\sum_{i=1}^N \theta_{\mu}(x_i+\epsilon_p).
\end{equation}
Normalizing the prices $x_i$ to lie in the interval $[0,1]$, the error of this approximation is

\begin{equation}
	|P' - P| = \left| \frac{1}{N}\sum_{i=1}^N \theta_{\mu}(x_i+\epsilon_p) - \int_0^1 \theta_{\mu}(x)p(x)dx\right| \le \epsilon_{\theta} + \epsilon_S \nonumber
\end{equation}
\begin{equation}
	\label{eqn:classical_error}
	\textrm{with }\quad \epsilon_{\theta} \equiv \left| \frac{1}{N}\sum_{i=1}^N \theta_{\mu}(x_i+\epsilon_p) - \frac{1}{N}\sum_{i=1}^N \theta_{\mu}(x_i)\right|, \quad \epsilon_S \equiv \left| \frac{1}{N}\sum_{i=1}^N \theta_{\mu}(x_i) - \int_0^1 \theta_{\mu}(x)p(x)dx\right|,
\end{equation}
where $\epsilon_{\theta}$ is the error stemming from the evaluation of the threshold function using an approximate derivative price, and $\epsilon_{S}$ the error from approximating the integral with Monte Carlo sampling.
We can derive the asymptotic dependence of $\epsilon_{\theta}$ on $\epsilon_p$ by considering the expression for $\epsilon_{\theta}$ in Eq.~\eqref{eqn:classical_error} in the continuous limit

\begin{eqnarray}
	\label{eqn:epsilon_theta_approx}
	\epsilon_{\theta} &=& \left| \int_0^1 \theta_{\mu}(x+\epsilon_p)p(x)dx - \int_0^1 \theta_{\mu}(x)p(x)dx\right| \nonumber\\
	&=& \left|\int_0^{\mu} p(x)dx + \int_{\mu}^{\mu+\epsilon_p} p(x)dx - \int_0^{\mu} p(x)dx \right| = \left|\int_{\mu}^{\mu+\epsilon_p} p(x)dx \right| \approx p(\mu)\epsilon_p + O(\epsilon_p^2),
\end{eqnarray}
where the last approximation follows by considering the left Riemann sum approximation to the integral \footnote{We can also consider the pricing error contribution to the threshold function in Eq.~\eqref{eqn:epsilon_theta_approx} as $\theta_{\mu}(x-\epsilon_p)$ and the same asymptotic dependence can be derived by considering the right Riemann sum approximation to the integral.}.
It then follows that $\epsilon_{\theta} = \bigO{\epsilon_p}$ with the constant factor depending on the shape of the probability distribution $p(x)$ at $\mu$.
Due to the two nested Monte Carlo evaluations, the complexity of this method scales as $\bigO{\epsilon_S^{-2}\epsilon_p^{-2}}$.

The \emph{semi-classical} approach proceeds similarly as in the purely classical case, the only difference being that the derivative price for each scenario is evaluated on a quantum computer using amplitude estimation.
In this case, the approximation error is the same, but the quadratic speedup of amplitude estimation makes the approximation to $P[X \le \mu]$ scale as $\bigO{\epsilon_S^{-2}\epsilon_p^{-1}}$.

\subsection{QAE \var{}}
\label{sec:qae_error}
With the QAE \var{} method we create a superposition over scenarios $\sum_i\sqrt{p(s_i)}\ket{s_i}$ with the operator of Eq.~\eqref{eqn:S_operator}, evaluate the derivative price $x_i$ for each scenario $s_i$ in superposition using amplitude estimation, and after applying a binary comparator at $\mu$, we read out the approximate probability $P'[X \le \mu]$ using another round of QAE (see Fig.~\ref{fig:qae_var_circuit}).
Contrary to the classical and semi-classical methods, in this case we are not limited to classical sampling of scenarios if we have a way to load the superposition of scenarios using a closed-form expression for $p(s_i)$.
As such, the approximation $P'[X\le\mu]$ in this case can be written as

\begin{equation}
	P'[X \le \mu] = \sum_{i=1}^N \theta_{\mu}(x_i + \epsilon_p)p(x_i)\delta x + \epsqae,
\end{equation}
where $\epsilon_p$ is the approximation error of using amplitude estimation to approximate the derivative price $x_i$ and $\epsqae$ is the error from the outer amplitude estimation.
The the sum over $x_i$ can be a (multi-dimensional) Riemann sum approximating an integral if $p(x)$ can directly loaded into the superposition of scenarios.
If the scenarios are instead sampled from $p(x)$, the above expression becomes a Monte Carlo approximation with $p(x_i)\delta x=1/N$.
In this case the error of the approximation can be written as

\begin{equation}
	|P' - P| = \left| \sum_{i=1}^N \theta_{\mu}(x_i + \epsilon_p)p(x_i)\delta x  - \int_0^1 \theta_{\mu}(x)p(x)dx\right| + \epsqae \le \epsilon_{\theta} + \epsilon_S + \epsqae \nonumber
\end{equation}
\begin{equation}
	\label{eqn:qae_error_terms}
	\textrm{where} \quad \epsilon_{\theta} \equiv \left| \sum_{i=1}^N \theta_{\mu}(x_i + \epsilon_p)p(x_i)\delta x  - \sum_{i=1}^N \theta_{\mu}(x_i)p(x_i)\delta x \right|, \quad \epsilon_S \equiv \left| \sum_{i=1}^N \theta_{\mu}(x_i)p(x_i)\delta x  - \int_0^1 \theta_{\mu}(x)p(x)dx\right|,
\end{equation}
and the asymptotic dependence of $\epsilon_{\theta}$ on $\epsilon_p$ follows similarly to the classical case, with $\epsilon_{\theta} = \bigO{\epsilon_p}$.

If the cost of loading the scenario superposition is $C(\epsilon_S)$, the total complexity of estimating $P[X \le \mu]$ using QAE \var{} is $\bigO{(C(\epsilon_S)+\epsilon_p^{-1})\epsilon_A^{-1}}$.
When the individual scenarios in the superposition are selected by sampling an appropriate probability distribution similarly to the classical and semi-classical approach, a sampling error $\epsilon_S$ requires $\epsilon_S=\bigO{1/\sqrt{N}}$ scenarios.
As this superposition can be loaded in $\bigO{\log N}$ depth\footnote{It is important to highlight that while this superposition can generally be loaded in $\bigO{\log N}$ depth, in practice, this complexity might include very large constant factors, as discussed in \cite{clader2022quantum}.} \cite{clader2022quantum}, the total complexity becomes $\bigO{(\log(1/\epsilon_S^2)+\epsilon_p^{-1})\epsilon_A^{-1}} = \tilde{\mathcal{O}}(\epsilon_p^{-1}\epsilon_A^{-1})$, where the tilde notation ignores poly-logarithmic terms.

\subsection{QSP \var{}}
\label{sec:qsp_error}
\begin{figure}[t]
  \centering
  \includegraphics[width=0.6\linewidth]{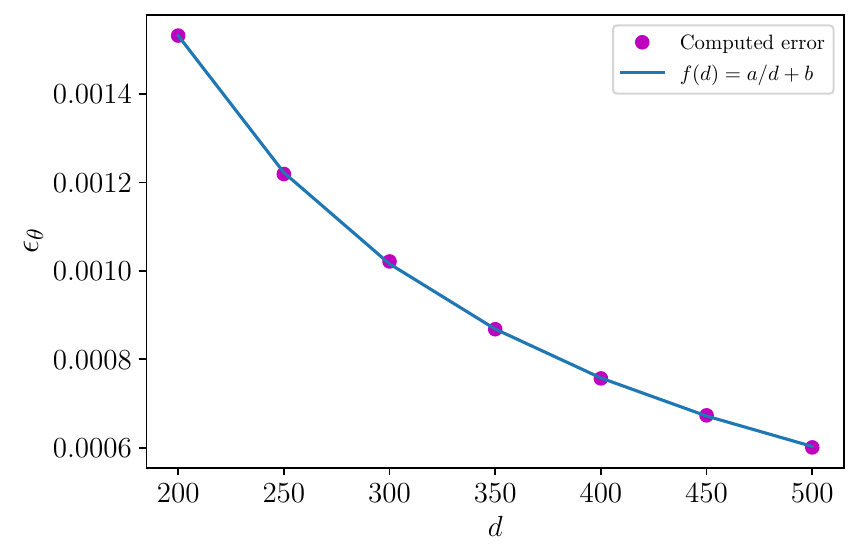}
  \caption{The approximation error $\epsilon_{\theta}$ in the estimation of $P[X \le \mu]$ due to the polynomial approximation used in the QSP \var{} method as shown in Eq.~\eqref{eqn:qsp_error_terms}. We pick the gap parameter $\Delta=10^{-3}$ and numerically compute the approximation error of $d$-degree polynomials generated using the optimization-based method described in Sec.~\ref{sec:thres_approximation}. This error is expected to scale as $\bigO{1/d}$ and fitting a function of the form $f(d) = a/d+b$ to the data in this case gives us $a \approx 0.31$.}
  \label{fig:qsp_approximation_error}
\end{figure}

The QSP \var{} algorithm follows similar steps as the QAE variant, but in this case the derivative prices are computed with the accuracy allowed of the $\mathcal{A}$ operator from Eq.~\eqref{eqn:A_operator} and do not incur an approximation error from amplitude estimation.
While there is still approximation error in the implementation of $\mathcal{A}$, it can be made exponentially small by increasing the number of qubits in the construction of the operator with only logarithmically increasing the oracle complexity of the algorithm \cite{chakrabarti2021threshold} and as such we ignore it in this error analysis.
On the other hand, because in this case we use QSP to approximate the threshold function of Eq.~\eqref{eqn:threshold_function}, we do not implement exactly the threshold function $\theta_{\mu}(x)$ but rather an approximation, which we denote as $\theta_{\mu}'(x)$.
As in the case of QAE \var{}, the preparation of the scenarios is not limited to sampling if there is a method to load the scenario superposition efficiently, so we use the same notation as the previous section to write the approximation to $P[X \le \mu]$ in this case as

\begin{equation}
	P'[X \le \mu] = \sum_{i=1}^N \theta_{\mu}'(x_i)p(x_i)\delta x + \epsqae.
\end{equation}
The approximation error is then given by

\begin{equation}
	|P' - P| = \left| \sum_{i=1}^N \theta_{\mu}'(x_i)p(x_i)\delta x - \int_0^1 \theta_{\mu}(x)p(x)dx\right| + \epsqae \le \epsilon_{\theta} + \epsilon_S + \epsqae \nonumber
\end{equation}
\begin{equation}
	\label{eqn:qsp_error_terms}
	\textrm{where} \quad \epsilon_{\theta} \equiv \left| \sum_{i=1}^N \theta_{\mu}'(x_i)p(x_i)\delta x - \sum_{i=1}^N \theta_{\mu}(x_i)p(x_i)\delta x\right|, \quad \epsilon_S \equiv \left| \sum_{i=1}^N \theta_{\mu}(x_i)p(x_i)\delta x - \int_0^1 \theta_{\mu}(x)p(x)dx\right|.
\end{equation}
Contrary to the classical, semi-classical and QAE cases, QSP allows us to bound the error from the first term $\epsilon_{\theta}$.
For a choice of the gap parameter $\Delta$, picking a polynomial degree of $d=\bigO{1/\Delta}$ will guarantee that $\epsilon_{\theta} \le \bigO{\Delta} \sim \bigO{1/d}$ \cite{low2017hamiltonian, dong2022ground}.
In Fig.~\ref{fig:qsp_approximation_error} we show numerically how this approximation error scales with the degree of the polynomial used to approximate $\theta_{\mu}'(x)$ for $\Delta=10^{-3}$ when the polynomial is generated using the method described in Sec.~\ref{sec:thres_approximation}.
The complexity of estimating $P[X\le \mu]$ then scales as $\bigO{(C(\epsilon_S) + \epsilon_{\theta}^{-1})\epsilon_A^{-1}}$, where again $C(\epsilon_S)$ denotes the cost of loading the superposition over scenarios.
In the case where the scenarios are selected by sampling a distribution and loaded in superposition in $\mathcal{O}(\log(N))$ depth, we get $\tilde{\mathcal{O}}\left(\epsilon_{\theta}^{-1}\epsilon_A^{-1} \right)$.

From the analysis in this section, we see that for a target error $\epsilon$ in the estimation of the probability in Eq.~\eqref{eqn:threshold_integral}, classical nested Monte Carlo scales as $\bigO{\epsilon^{-4}}$ and the semi-classical method scales as $\bigO{\epsilon^{-3}}$, where the advantage stems from the quadratic advantage of QAE compared to classical sampling in the inner (pricing) Monte Carlo step.
The semi-classical method cannot provide further advantage in the outer Monte Carlo sampling step, as both of these methods are limited to sampling scenarios from an appropriate probability distribution.
When also restricted to scenario sampling, the QAE and QSP-based quantum methods both scale as $\tilde{\mathcal{O}}(\epsilon^{-2})$.
If the relevant probability distribution of scenarios can be loaded efficiently in superposition, we do not have to load individually sampled scenarios in superposition, and the error $\epsilon_S$ in Eq.~\eqref{eqn:qae_error_terms} and Eq.~\eqref{eqn:qsp_error_terms} can be made exponentially small, which is not possible with the classical and semi-classical approaches.

\section{Comparison of the Quantum Methods for Estimating \var{}}
\label{sec:compare_methods}
The QAE and QSP \var{} quantum estimation methods described in the previous sections are based on the same procedure and it is not obvious whether one outperforms the other in terms of required quantum resources for a target estimation accuracy.
The only difference between the two methods is how the probability that a portfolio value under a number of scenarios is smaller than a \var{} candidate $\mu$ is encoded in an amplitude so that we can perform QAE to read it out.
The QAE method performs an analog$\rightarrow$digital$\rightarrow$analog transformation whereby the portfolio values initially encoded as amplitudes, are digitized approximately into a quantum register by the inner round of QAE before being transformed appropriately back into amplitudes using a binary comparator circuit.
On the other hand, the QSP framework allows us to perform the necessary transformation to the amplitudes without the need for an intermediate binary representation which removes that source of error in the process.
It does however introduce an error from the approximate implementation of the threshold function which encodes the value to be extracted with QAE.
Additionally, looking at the quantum circuits for each method in Fig.~\ref{fig:qae_var_circuit} and Fig.~\ref{fig:qsp_var_circuit}, the QSP circuit is simpler in that a) it only requires one additional qubit other than those required to implement the $\mathcal{S}$ and $\mathcal{A}$ operators which load the scenario superposition and encode the derivative price into an amplitude respectively, b) the $\mathcal{A}$ operator is only controlled on the scenario qubits while it is controlled on one more qubit in the QAE case by virtue of the controlled invocation $\mathcal{Q}$ operator, and c) while the QSP method contains single-qubit phase rotations, the $\mathcal{Q}$ operators in the QAE method require reflections.

The total resources required for both methods can be summarized as

\begin{equation}
	\label{eqn:total_cost}
	C_{\textrm{\var{}}} = C_{S}C_{\mu}C_{\textrm{QAE}}\cdot k,
\end{equation}
where $C_{S}$ is the cost of implementing the scenario superposition unitary $\mathcal{S}$, $C_{\mu}$ denotes the resources required by either method to generate a state where the probability of measuring a qubit in the $\ket{0}$ state gives the probability that the derivative portfolio value falls below a value $\mu$, $C_{\textrm{QAE}}$ is the cost of QAE to read out that probability and $k$ is the number of bisection search iterations required to compute the \var{} estimate to a desired accuracy.
The cost component $C_{\mu}$ is the main difference between the two methods in terms of the overall algorithmic cost so we focus on estimating the cost of this component in each case.
Specifically, we measure the cost by the number of oracle invocations to the pricing operator $\mathcal{A}$ (and its inverse $\mathcal{A}^{\dagger}$) which is the most expensive component in derivative pricing \cite{chakrabarti2021threshold}.
The QAE method requires one invocation of $\mathcal{A}$ to set up the initial state Eq.~\eqref{eqn:AS_operator} and $m$ ancilla qubits to discretize the value of the derivative portfolio under the scenarios.
Each ancilla qubit $j \in [0, m-1]$ controls $2^j$ invocations of the $\mathcal{Q}$ operator as shown in Fig.~\ref{fig:qae_var_circuit}, and $\mathcal{Q}$ is comprised of one instance of $\mathcal{A}$ and one of $\mathcal{A}^{\dagger}$, giving a total of $\nqae(m) = 2^m+1$ oracle calls.
The QSP method uses a $d$-degree polynomial to approximate the threshold function $\theta_{\mu}(x)$ of Eq.~\eqref{eqn:threshold_function} and the its implementation requires $\nqsp(d) = d-1$ invocations of $\mathcal{A}$ as shown in Eq.~\eqref{eqn:QSP_general_transformation} and Fig.~\ref{fig:qsp_var_circuit}.

We estimate the values of $\nqae$ and $\nqsp$ numerically by sampling simulated scenario prices from a normal distribution with mean $0.5$ and standard deviation $0.09$ and look for the \var{} at the $\alpha=99\%$ level.
The mean and standard deviation of the normal distribution were chosen so that the sampled values lie with high probability in the interval $[0, 1]$ and the $\alpha=99\%$ confidence level is typical for \var{} estimations in practice \cite{OBRIEN2017215}.
We simulate the QAE \var{} estimation algorithm as described in Sec.~\ref{sec:qae_var}, where for each sampled scenario price we generate the superposition of binary strings returned by QAE as possible approximations to the value, weighed by the corresponding probabilities of Eq.~\eqref{eqn:qae_probabilities}.
Starting with $\mu=0.5$, we calculate the aggregate probability $P_0$ across all scenarios that the price is smaller than $\mu$.
Then, for a parameter $\epsilon_A$ we determine bounds $[p_l, p_h] = [P_0-\epsilon_A, P_0+\epsilon_A]$ as the confidence interval given by an amplitude estimation round to determine $P_0$ with target accuracy $\epsilon_A$.
We run a bisection search over $\mu$ until we find that $1-\alpha=0.01 \in [p_l, p_h]$ and we return that value of $\mu$ as the \var{} estimate.

The QSP method is simulated similarly, the only difference being how we calculate $P_0$ for a given value of $\mu$.
In this case, we pick a value for the polynomial degree $d$ and for each value of $\mu$ during bisection search we generate a $d$-degree polynomial approximation to the threshold function $\theta_{\mu}(x)$ using the method described in Sec.~\ref{sec:thres_approximation} and apply the resulting function to the sampled scenario prices.
Subsequently, we generate the same confidence interval for the amplitude estimation round given a target accuracy $\epsilon_A$ as in the QAE method, and impose the same stopping condition.
The equivalent classical estimate is computed by sorting the sampled scenario prices and finding the value at the $1-\alpha$ percentile.
Because the simulations of the quantum methods use sampled scenarios, the classical estimate represents the lowest bound either quantum method can achieve.

\begin{figure}[t]
  \centering
  \includegraphics[width=0.6\linewidth]{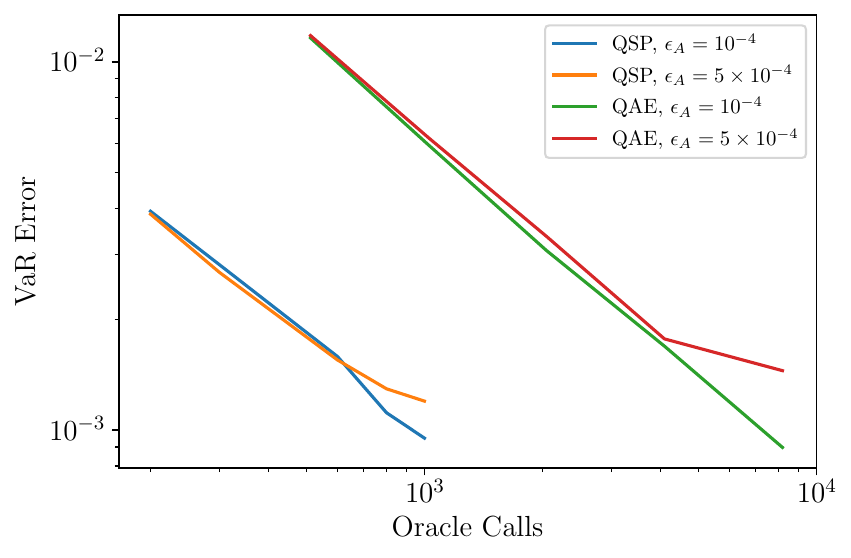}
  \caption{\var{} estimation errors as a function of the number of oracle calls for the QSP and QAE quantum methods. For each data point we generate $100$ samples, and for each sample we randomly generate $N=5k$ scenario prices from the normal probability distribution $\mathcal{N}(\mu=0.5, \sigma^2=0.09^2)$ and compute the \var{} estimate by simulating the QAE and QSP algorithms as described in Sec.~\ref{sec:compare_methods}. The error of each sample is calculated as the absolute value of the difference between the quantum \var{} estimate and the classical \var{} estimate from the same generated scenario prices. We simulate each algorithm for two values of the final amplitude estimation accuracy $\epsilon_A$ and plot the average error across all samples for that each data point. On average, we observe that the QSP method requires 10x fewer oracle calls than QAE for the same \var{} error.
  While increasing number of oracle calls $N_o$ leads to the \var{} error decreasing as $\mathcal{O}(N_o^{-1})$, notice that this scaling starts to depart from linearity as the amplitude estimation accuracy $\epsilon_A$ gets close to the accuracy $1/N_o$ with which the amplitude is encoded.
  Eventually, the scaling plateaus as the increased encoding accuracy becomes too fine to be resolved by an amplitude estimation process of fixed accuracy.}

  \label{fig:sampled_var_error_qsp_ae}
\end{figure}

The results from the simulation for both QAE and QSP methods for different values of oracle call invocations and the final amplitude estimation accuracy $\epsilon_A$ are shown in Fig.~\ref{fig:sampled_var_error_qsp_ae}.
Each data point is comprised of $100$ independent samples, where for each sample we randomly generate $N=5k$ scenario prices from the normal probability distribution $\mathcal{N}(\mu=0.5, \sigma^2=0.09^2)$, simulate the QAE and QSP algorithms and then compute the error of each sample as the absolute value of the difference between the quantum \var{} estimate and the classical \var{} estimate from the same generated scenario prices.
In the figure we plot the average error across all samples for each data point.
One initial observation is that the QSP method performs consistently better for the error range examined, requiring on average 10x fewer oracle calls than QAE for the same target error.
In addition, we notice that as the number of oracle calls increase, the probability $P'$ we are estimating each time using amplitude estimation is encoded with increasing accuracy, but clearly at some point this increased encoding accuracy will fail to reduce the \var{} estimation error if the value of $\epsilon_A$ is not small enough to resolve the difference.
This is what we observe at the tail end of the plots in Fig.~\ref{fig:sampled_var_error_qsp_ae}, where the \var{} error in the $\epsilon_A=5\times 10^{-4}$ plots starts to decrease more slowly than the $\epsilon_A= 10^{-4}$ plots.
Until that point, we calculate that for a fixed value of $\epsilon_A$, the \var{} error decreases approximately linearly as the number of oracle calls increases for both QAE and QSP cases (the slope of a log-log fit is approximately $-0.9$ for all four plots), as expected from the error analysis in Sec.~\ref{sec:qae_error} and Sec.~\ref{sec:qsp_error} respectively where both errors scale as $\mathcal{O}(\epsilon^{-1}_p)$ with everything else kept constant.

Our choice of parameters $N$, $\mu$ and $\sigma$ was made to simplify the numerical simulations, but we observe qualitatively similar results for different choices of distribution parameters $\mu$ and $\sigma$ as well as the number of scenario prices generated for each data point.

\section{Conditions for Quantum Advantage}
\label{sec:quantum_advantage}

On account of the numerical results in the previous section indicating that the QSP-based \var{} method outperforms the corresponding QAE variant, in this section we turn our attention on how the \var{} estimation with QSP fares against classical methods and what conditions (if any) there are for potential quantum advantage.
As discussed in Sec.~\ref{sec:error_analysis}, the estimation error of the QSP method is lower-bounded by the classical sampling error of $\bigO{\epsilon^{-2}}$ when the superposition of scenarios is generated with samples from a probability distribution.
Assuming no particular structure in the representation of the sampled scenarios, the superposition of $N$ scenarios can be loaded to accuracy $\epsilon$ using the QRAM-based state preparation routine from Ref.~\cite{clader2022quantum} with T-depth $\bigO{\log(N) +
\log(1/\epsilon)}$, at a cost of $\bigO{N}$ T-count.
On the other hand, if the scenarios can be prepared by discretizing a multivariate probability distribution of the relevant risk factors and loaded efficiently in superposition, the lower bound on the \var{} estimation error can be made exponentially small by increasing the number of qubits representing the distribution.
In the case that the probability distribution can be modeled explicitly with an analytic density function, the scenario superposition could be prepared using the re-parameterization method from Ref.~\cite{chakrabarti2021threshold}.
Alternatively, if the scenario probabilities have to be inferred from historical data, quantum representations of financially relevant copulas \cite{milek2020quantum, zhu2022copula} could potentially be employed if they can scale to scenarios of high dimensionality.
In either case, the resources required to generate the scenario superposition will be highly dependent on the nature of the target probability distribution if explicit loading is possible, and the number of risk factors that make up each scenario as well as the required granularity for each risk factor (as shown in Eq.~\eqref{eqn:scenario_encoding}).
As such, in this section we do not examine specific choices of scenario preparation methods, and instead consider the cost of loading the superposition to be a free variable which needs to be taken into account when estimating the total resources for the algorithm.

As a benchmark to examine the conditions for practical quantum advantage, we consider the autocallable derivative product studied in Ref.~\cite{chakrabarti2021threshold}.
We calculate the resources required by the QSP-based \var{} algorithm to estimate the \var{} of the product to the same accuracy as a classical Monte Carlo method, assuming that the product is evaluated under scenarios such that the resulting prices are distributed normally.
We use the latest available resource estimates from Ref.~\cite{stamatopoulos2023derivative}, whereby an autocallable derivative product can be priced to within $\epsilon_p=2\times 10^{-3}$ in one second classically and the $\mathcal{A}$ operator required to price the contract on a quantum computer can be constructed with a T-depth of $T_A = 3900$.

The pricing of this product under $N$ scenarios is simulated classically by randomly generating $N$ values from a normal distribution $\mathcal{N}(\mu, \sigma^2)$ representing $N$ scenario prices, and adding a noise term sampled from $\mathcal{N}(0, \epsilon_p^2)$.
Then we compute $\varvalue$ from Eq.~\eqref{eqn:var_definition} by sorting the resulting simulated prices from smallest to largest, and picking the value at the $1-\alpha$ percentile.
The error of the \var{} estimate from the classical simulation can be calculated as $|\varvalue - \Phi^{-1}(1-\alpha)|$, where $\Phi(x)$ is the cumulative distribution function of the distribution $\mathcal{N}(\mu, \sigma^2)$.
We repeat this process $200$ times, and define the error of the classical algorithm $\epsilon_C$ as the standard deviation of the resulting distribution of errors.
Since we assume the contract can be priced classically in one second, for $N$ scenarios, we consider the runtime of the end-to-end classical \var{} estimation to be $N$ seconds.

To evaluate the performance of the quantum QSP \var{} algorithm, we similarly generate $N$ values from the normal distribution $\mathcal{N}(\mu, \sigma^2)$ representing $N$ scenario prices and simulate the QSP algorithm as described in Algorithm~\ref{algo:qsp_var}.
The error of the quantum estimate is given by $|\mu_{\alpha}^2 - \Phi^{-1}(1-\alpha)|$, where $\mu_{\alpha}$ is the estimate of $\sqrt{\varvalue}$ returned by the QSP algorithm.
Repeating the process $200$ times, we define the error of the quantum algorithm $\epsilon_Q$ as the value at the $68$th percentile of the resulting distribution of errors.
This way we can say that with probability $68\%$ the estimation error of the algorithm does not exceed $\epsilon_Q$, which corresponds to the same confidence in the estimate of the classical algorithm, defined as the error at one standard deviation.

We then proceed to estimate the T-depth of the end-to-end process, as a proxy for the total runtime.
The T-depth of each iteration, which is the T-depth of the circuit of Fig.~\ref{fig:qsp_var_circuit} is given by

\begin{equation}
	T_i = T_S + dT_A + dT_R,
\end{equation}
where $T_S$ is the T-depth of the scenario preparation unitary $\mathcal{S}$, $T_A=3900$ is the T-depth of the $\mathcal{A}$ operator \cite{stamatopoulos2023derivative}, $d$ is the polynomial degree used for the approximation to the threshold function and $T_R$ is the T-depth of implementing each controlled $R_z$ rotation.
Using the method in \cite{ross_2016}, the $R_z$ rotation can be performed to precision $\epsilon_R$ with a T-depth of approximately $3\log_2(1/\epsilon_R)$ and the decomposition in \cite{Kim_Efficient_2018} allows us to perform the controlled rotation with an $R_z$ depth of one, using one ancilla qubit.
We measure the cost of each round of QAE for target accuracy $\epsilon_A > 0$ and confidence level $1-\alpha_k, \alpha_k \in (0,1)$ by the bound derived in \cite{grinko2021iterative}

\begin{equation}
\label{eqn:AE_oracle_calls}
N_{\text{oracle}}^{\text{wc}}
\leq \frac{1.4}{\epsilon_A}\log\left(\frac{2}{\alpha_k} \log_2\left(\frac{\pi}{4\epsilon_A}\right)\right),
\end{equation}
where $N_{\text{oracle}}^{\text{wc}}$ denotes the worst-case number of QAE oracle calls.
In our case, the QAE oracle call consists of one invocation of the entire circuit in Fig.~\ref{fig:qsp_var_circuit} and one to its inverse (where we ignore the cost of reflections in the QAE oracle).
Thus, the end-to-end T-depth of the QSP \var{} algorithm which goes through $k$ rounds of QAE is given by

\begin{equation}
	\label{eqn:t_depth_qsp}
	T_{\textrm{QSP}} = \frac{2.8k}{\epsilon_A}\log\left(\frac{2}{\alpha_k} \log_2\left(\frac{\pi}{4\epsilon_A}\right)\right) \left(T_S + dT_A + 3d\log_2(1/\epsilon_R) \right).
\end{equation}
Additionally, we choose $\epsilon_R$ such that the total error from all the $R_z$ rotations is the same order of magnitude as the error from the polynomial approximation to the threshold function, which is $\gtrsim 10^{-4}$ as shown in Fig.~\ref{fig:qsp_approximation_error}.
In our numerical simulations the polynomial degrees we use satisfy $d\le 1000$, so we pick $\epsilon_R=10^{-7}$.

The remaining free parameters in the total cost of the QSP algorithm are the scenario loading cost $T_S$, the polynomial degree $d$ used for QSP and the target accuracy $\epsilon_A$ with which we estimate the encoded probability in each round of QAE.
Additionally, for a fixed number of scenarios $N$, the \var{} estimation error is only dependent on the choices of $d$ and $\epsilon_A$.
In order to benchmark the performance of the QSP algorithm, for a choice of $N$, we numerically look for the optimal values of parameters $(d, \epsilon_A)$ such that the estimation error of the quantum algorithm $\epsilon_Q$ is equal to that of the classical algorithm $\epsilon_C$ and calculate the overall T-depth given by Eq.~\eqref{eqn:t_depth_qsp} with $T_S$ left as the only remaining free cost parameter.
Once we estimate the optimal values of $(d, \epsilon_A)$, we can additionally define the minimum rate at which that a quantum processor would need to execute T-gates so that the physical runtime of the QSP method would equal the physical runtime of a classical processor.
That \emph{logical} QPU clock rate is defined as $L \equiv T_{\textrm{QSP}}/N$, where, as described above, we assume it takes $N$ seconds to classically price $N$ scenarios.

We use $\mu=0.5$ and $\sigma=0.09$ as the parameters of the normal distribution from which we draw scenario samples, and simulate the classical \var{} estimation method for $\alpha=99\%$.
The QSP algorithm is simulated for the same value of $\sigma$, but in this case we average the results over $\mu \in [0.45, 0.48, 0.5, 0.52, 0.55]$.
The reason for this is that the QSP method can do better (or worse) depending on where the true \var{} value lies, because of the bisection search performed.
Specifically, the estimation error of the quantum method will depend on the distance between the true \var{} value and the values reachable by bisection search in the interval $[0,1]$.
We therefore use this averaging over values of $\mu$ to benchmark the average performance of the algorithm.
On the other hand, the estimation error has no dependence on the true value of \var{} classically, so we fix the value of $\mu$ in that case.
We chose $\sigma=0.09$ for the simulated scenario distribution such that the sampled scenarios are in the interval $[0,1]$ with high probability.
We have also simulated the algorithms for $\sigma \in [0.05, 0.07]$ and our results remain qualitatively unchanged.

\begin{figure}[t]
  \centering
  \includegraphics[width=0.6\linewidth]{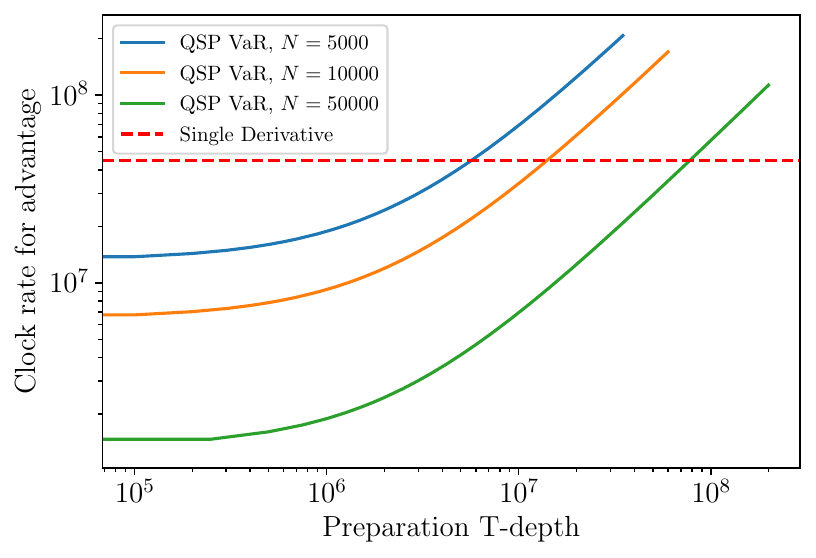}
  \caption{Minimum QPU logical clock rate required for the QSP \var{} method to match classical methods in total execution time as a function of the scenario loading T-depth. For reference, the horizontal dashed line indicates the minimum logical clock rate required for quantum advantage in the pricing of a single derivative as estimated in Ref.~\cite{stamatopoulos2023derivative}. Both classical and quantum algorithms are simulated by sampling scenarios from a normal distribution as described in Sec.~\ref{sec:quantum_advantage}, and for each choice of number of scenarios $N$, we search for the parameters $(d, \epsilon_A)$ (controlling the polynomial approximation to the threshold function and the accuracy of QAE respectively) such that the estimation error from the QSP algorithm matches that of the classical algorithm. For $N=50$k, we see that there is a possibility for reduction of the logical clock rate required for quantum advantage for up to a factor $\sim 30$x compared to that required in the pricing of a single derivative. For all choices of $N$ shown here, we numerically estimate that the QSP method requires at least $d=600$ and $\epsilon_A=1.2\times 10^{-3}$ in order to match the corresponding estimation error of the classical method.}
  \label{fig:clock_rate_advantage}
\end{figure}

In Fig.~\ref{fig:clock_rate_advantage} we show the logical clock rate required for quantum advantage as estimated from the numerical simulations, as function of the scenario preparation T-depth $T_S$, for different numbers of scenarios $N$.
For reference, we also show the logical clock rate (45Mhz) required for quantum advantage in pricing a single derivative calculated in Ref.~\cite{stamatopoulos2023derivative}.
Because the runtime of the classical and semi-classical methods scales linearly with $N$, as $N$ increases, the QSP method starts to provide room for quantum advantage as long as the scenario preparation can be prepared with a smaller T-depth than a certain threshold, which can be derived from Eq.~\eqref{eqn:t_depth_qsp}.
For small preparation T-depths, the value of $T_S$ is dominated by the resources required for QSP, reducing the logical clock rate for advantage until the two terms become comparable.
For $N=50$k, we notice that there is an advantage over single derivative pricing as long as the scenario superposition can be prepared with a T-depth $T_S \lesssim 8 \times 10^{7}$.
In the case that the superposition can be prepared with a T-depth $T_S \lesssim 3 \times 10^{5}$, the logical clock rate required for advantage is reduced by a factor of $\sim 30$x.

We note here that these estimates for quantum advantage are based on the assumption that the superposition over scenarios is created from sampled scenarios.
In the case where an explicit probability distribution of scenarios can be loaded efficiently as a discretized probability density function, the error from the scenario sampling would decrease exponentially and allow the possibility of further quantum advantage.

\section{Discussion}
\label{sec:discussion}
While the efficacy of quantum computers in derivative pricing has been studied in the context of estimating the value of individual derivatives, in various relevant contexts, the quantity of interest is not the individual derivative price itself, but rather metrics which characterize the risk profile of financial derivatives.
Because calculating these metrics often requires pricing derivatives repeatedly, the power of quantum computers could be harnessed more effectively by considering the aggregate computation instead of the application of quantum computing to the individual components.
As such, devising quantum algorithms which use the oracles constructed for the pricing of derivatives as subcomponents, opens up a new avenue for quantum advantage in the context of derivative pricing \cite{Stamatopoulos_2022}.
In this work, we consider the problem of estimating the Value at Risk and the Conditional Value at Risk of financial derivatives.
We first extend the algorithm introduced in Ref.~\cite{Woerner_2019, egger2019credit} to estimate the \var{} of financial derivatives and then introduce a quantum algorithm based on the Quantum Signal Processing framework to estimate these metrics.
Even though we show that asymptotically these two algorithms have similar performance, we show that the use of QSP in practice allows for a more efficient algorithm, a conclusion which has also been observed in different contexts \cite{martyn2021efficient, Lin_2020, Rall_2023}.
This result also highlights the practical advantage of QSP-based methods which not only allow the application of generic transformations to appropriately encoded values of interest, but that they do so with minimal overhead.
While our formulation of derivative pricing in the QSP framework was used in this case to calculate the \var{} and \cvar{} risk metrics, we expect that it will open up new research avenues in the larger context of derivative pricing using quantum computers.

Additionally, we study the possibility of quantum advantage by using this algorithm over equivalent classical methods.
The performance of the algorithm relies on the ability to prepare a superposition of scenarios efficiently and different approaches to that end have been proposed \cite{chakrabarti2021threshold, milek2020quantum, zhu2022copula}.
During the writeup of this manuscript, a state preparation method was introduced to load a uniform superposition of subsets of computational basis states scaling as $\bigO{\log_2 M}$ for $M$ states in superposition, with low overhead \cite{Shukla_2024}.
The superposition over scenarios encoded using Eq.~\eqref{eqn:scenario_encoding} is precisely a question of generating a uniform superposition over subsets of computational basis states, and therefore determining the applicability of this method to the \var{} algorithms we describe here is a very promising research direction.
In this work, we have left the scenario preparation as an open question and instead, through numerical simulations, we estimate upper bounds on the resources required to prepare the necessary superposition in order for quantum advantage to exist.
We find that in certain settings, the quantum algorithm can provide a reduction in the logical clock rate required for quantum advantage estimated in Ref.~\cite{stamatopoulos2023derivative} by $\sim 30$x.
While this estimate is based on pricing the derivative under scenarios sampled from a relevant probability distribution, quantum computing allows the possibility of using the full (discretized) multivariate probability distribution in the calculation, an option not available to classical methods due to the curse of dimensionality.
An extension of this work would be to study which (relevant) probability distributions can be loaded efficiently on a quantum computer and the impact on the estimate of possible quantum advantage calculated in this work.

In our comparison between the performance of quantum and classical methods, we have assumed that classical methods for the \var{} estimation of financial derivatives require pricing derivative contracts across a fixed number of scenarios $N$, such that the total complexity scales as $\bigO{N/\epsilon^{2}}$.
In Ref.~\cite{giles2019}, the authors argue that for sufficiently smooth probability density functions, an adaptive sampling technique combined with Multi-level Monte Carlo can reduce the complexity of general \var{} and \cvar{} estimation to $\tilde{\mathcal{O}}(\epsilon^{-2})$.
One core element in that approach is that the probability distribution behind the scenarios used in the \var{} estimation can be easily sampled, such that the number of samples generated can be adjusted to fit the target accuracy of the calculation.
However, in practice, it is extremely difficult to generate such probability distribution across (typically) thousands of market factors, and most commonly a fixed number of scenarios is created based on historical data that capture market features deemed important from a financial modeler's perspective.
That said, Multi-level Monte Carlo can also be applied in a quantum setting \cite{An_2021}, and adaptive samples can similarly be incorporated in the quantum process, by adjusting the accuracy with which the derivative price is encoded as a quantum amplitude.
We leave the detailed analysis of this approach to a future study.

\begin{acknowledgments}
We thank Bryce Fuller, Patrick Rall and Farrokh Labib for discussions regarding Quantum Signal Processing and feedback on this manuscript, Kunal Kishore for his technical and business insights regarding the \var{} estimation of financial derivatives, and Noelle Ibrahim for discussions on classical adaptive methods for estimating \var{}.
\end{acknowledgments}

\bibliographystyle{apsrev4-1_custom}
\bibliography{derivative_var}

\begin{thebibliography}{39}%
\makeatletter
\providecommand \@ifxundefined [1]{%
 \@ifx{#1\undefined}
}%
\providecommand \@ifnum [1]{%
 \ifnum #1\expandafter \@firstoftwo
 \else \expandafter \@secondoftwo
 \fi
}%
\providecommand \@ifx [1]{%
 \ifx #1\expandafter \@firstoftwo
 \else \expandafter \@secondoftwo
 \fi
}%
\providecommand \natexlab [1]{#1}%
\providecommand \enquote  [1]{``#1''}%
\providecommand \bibnamefont  [1]{#1}%
\providecommand \bibfnamefont [1]{#1}%
\providecommand \citenamefont [1]{#1}%
\providecommand \href@noop [0]{\@secondoftwo}%
\providecommand \href [0]{\begingroup \@sanitize@url \@href}%
\providecommand \@href[1]{\@@startlink{#1}\@@href}%
\providecommand \@@href[1]{\endgroup#1\@@endlink}%
\providecommand \@sanitize@url [0]{\catcode `\\12\catcode `\$12\catcode
  `\&12\catcode `\#12\catcode `\^12\catcode `\_12\catcode `\%12\relax}%
\providecommand \@@startlink[1]{}%
\providecommand \@@endlink[0]{}%
\providecommand \url  [0]{\begingroup\@sanitize@url \@url }%
\providecommand \@url [1]{\endgroup\@href {#1}{\urlprefix }}%
\providecommand \urlprefix  [0]{URL }%
\providecommand \Eprint [0]{\href }%
\providecommand \doibase [0]{http://dx.doi.org/}%
\providecommand \selectlanguage [0]{\@gobble}%
\providecommand \bibinfo  [0]{\@secondoftwo}%
\providecommand \bibfield  [0]{\@secondoftwo}%
\providecommand \translation [1]{[#1]}%
\providecommand \BibitemOpen [0]{}%
\providecommand \bibitemStop [0]{}%
\providecommand \bibitemNoStop [0]{.\EOS\space}%
\providecommand \EOS [0]{\spacefactor3000\relax}%
\providecommand \BibitemShut  [1]{\csname bibitem#1\endcsname}%
\let\auto@bib@innerbib\@empty
\bibitem [{\citenamefont {Montanaro}(2015)}]{montanaro2015quantum}%
  \BibitemOpen
  \bibfield  {author} {\bibinfo {author} {\bibfnamefont {A.}~\bibnamefont
  {Montanaro}},\ }\bibfield  {title} {\enquote {\bibinfo {title} {\emph
  {Quantum speedup of Monte Carlo methods}},}\ }\href
  {https://doi.org/10.1098/rspa.2015.0301} {\bibfield  {journal} {\bibinfo
  {journal} {Proceedings of the Royal Society of London A: Mathematical,
  Physical and Engineering Sciences}\ }\textbf {\bibinfo {volume} {471}}
  (\bibinfo {year} {2015})}\BibitemShut {NoStop}%
\bibitem [{\citenamefont {Rebentrost}\ \emph {et~al.}(2018)\citenamefont
  {Rebentrost}, \citenamefont {Gupt},\ and\ \citenamefont
  {Bromley}}]{rebentrost2018quantum}%
  \BibitemOpen
  \bibfield  {author} {\bibinfo {author} {\bibfnamefont {P.}~\bibnamefont
  {Rebentrost}}, \bibinfo {author} {\bibfnamefont {B.}~\bibnamefont {Gupt}}, \
  and\ \bibinfo {author} {\bibfnamefont {T.~R.}\ \bibnamefont {Bromley}},\
  }\bibfield  {title} {\enquote {\bibinfo {title} {\emph {Quantum computational
  finance: Monte Carlo pricing of financial derivatives}},}\ }\href
  {https://doi.org/10.1103/PhysRevA.98.022321} {\bibfield  {journal} {\bibinfo
  {journal} {Phys. Rev. A}\ }\textbf {\bibinfo {volume} {98}},\ \bibinfo
  {pages} {022321} (\bibinfo {year} {2018})}\BibitemShut {NoStop}%
\bibitem [{\citenamefont {Woerner}\ and\ \citenamefont
  {Egger}(2019)}]{Woerner_2019}%
  \BibitemOpen
  \bibfield  {author} {\bibinfo {author} {\bibfnamefont {S.}~\bibnamefont
  {Woerner}}\ and\ \bibinfo {author} {\bibfnamefont {D.~J.}\ \bibnamefont
  {Egger}},\ }\bibfield  {title} {\enquote {\bibinfo {title} {\emph {Quantum
  risk analysis}},}\ }\href {https://doi.org/10.1038/s41534-019-0130-6}
  {\bibfield  {journal} {\bibinfo  {journal} {npj Quantum Information}\
  }\textbf {\bibinfo {volume} {5}} (\bibinfo {year} {2019})}\BibitemShut
  {NoStop}%
\bibitem [{\citenamefont {Stamatopoulos}\ \emph {et~al.}(2020)\citenamefont
  {Stamatopoulos}, \citenamefont {Egger}, \citenamefont {Sun}, \citenamefont
  {Zoufal}, \citenamefont {Iten}, \citenamefont {Shen},\ and\ \citenamefont
  {Woerner}}]{Stamatopoulos_2020}%
  \BibitemOpen
  \bibfield  {author} {\bibinfo {author} {\bibfnamefont {N.}~\bibnamefont
  {Stamatopoulos}}, \bibinfo {author} {\bibfnamefont {D.~J.}\ \bibnamefont
  {Egger}}, \bibinfo {author} {\bibfnamefont {Y.}~\bibnamefont {Sun}}, \bibinfo
  {author} {\bibfnamefont {C.}~\bibnamefont {Zoufal}}, \bibinfo {author}
  {\bibfnamefont {R.}~\bibnamefont {Iten}}, \bibinfo {author} {\bibfnamefont
  {N.}~\bibnamefont {Shen}}, \ and\ \bibinfo {author} {\bibfnamefont
  {S.}~\bibnamefont {Woerner}},\ }\bibfield  {title} {\enquote {\bibinfo
  {title} {\emph {Option Pricing using Quantum Computers}},}\ }\href
  {https://doi.org/10.22331/q-2020-07-06-291} {\bibfield  {journal} {\bibinfo
  {journal} {Quantum}\ }\textbf {\bibinfo {volume} {4}},\ \bibinfo {pages}
  {291} (\bibinfo {year} {2020})}\BibitemShut {NoStop}%
\bibitem [{\citenamefont {Herbert}(2022)}]{Herbert_2022}%
  \BibitemOpen
  \bibfield  {author} {\bibinfo {author} {\bibfnamefont {S.}~\bibnamefont
  {Herbert}},\ }\bibfield  {title} {\enquote {\bibinfo {title} {\emph {Quantum
  Monte Carlo Integration: The Full Advantage in Minimal Circuit Depth}},}\
  }\href {https://doi.org/10.22331/q-2022-09-29-823} {\bibfield  {journal}
  {\bibinfo  {journal} {Quantum}\ }\textbf {\bibinfo {volume} {6}},\ \bibinfo
  {pages} {823} (\bibinfo {year} {2022})}\BibitemShut {NoStop}%
\bibitem [{\citenamefont {Chakrabarti}\ \emph {et~al.}(2021)\citenamefont
  {Chakrabarti}, \citenamefont {Krishnakumar}, \citenamefont {Mazzola},
  \citenamefont {Stamatopoulos}, \citenamefont {Woerner},\ and\ \citenamefont
  {Zeng}}]{chakrabarti2021threshold}%
  \BibitemOpen
  \bibfield  {author} {\bibinfo {author} {\bibfnamefont {S.}~\bibnamefont
  {Chakrabarti}}, \bibinfo {author} {\bibfnamefont {R.}~\bibnamefont
  {Krishnakumar}}, \bibinfo {author} {\bibfnamefont {G.}~\bibnamefont
  {Mazzola}}, \bibinfo {author} {\bibfnamefont {N.}~\bibnamefont
  {Stamatopoulos}}, \bibinfo {author} {\bibfnamefont {S.}~\bibnamefont
  {Woerner}}, \ and\ \bibinfo {author} {\bibfnamefont {W.~J.}\ \bibnamefont
  {Zeng}},\ }\bibfield  {title} {\enquote {\bibinfo {title} {\emph {A Threshold
  for Quantum Advantage in Derivative Pricing}},}\ }\href
  {https://doi.org/10.22331/q-2021-06-01-463} {\bibfield  {journal} {\bibinfo
  {journal} {Quantum}\ }\textbf {\bibinfo {volume} {5}},\ \bibinfo {pages}
  {463} (\bibinfo {year} {2021})}\BibitemShut {NoStop}%
\bibitem [{\citenamefont {Stamatopoulos}\ \emph {et~al.}(2022)\citenamefont
  {Stamatopoulos}, \citenamefont {Mazzola}, \citenamefont {Woerner},\ and\
  \citenamefont {Zeng}}]{Stamatopoulos_2022}%
  \BibitemOpen
  \bibfield  {author} {\bibinfo {author} {\bibfnamefont {N.}~\bibnamefont
  {Stamatopoulos}}, \bibinfo {author} {\bibfnamefont {G.}~\bibnamefont
  {Mazzola}}, \bibinfo {author} {\bibfnamefont {S.}~\bibnamefont {Woerner}}, \
  and\ \bibinfo {author} {\bibfnamefont {W.~J.}\ \bibnamefont {Zeng}},\
  }\bibfield  {title} {\enquote {\bibinfo {title} {\emph {Towards Quantum
  Advantage in Financial Market Risk using Quantum Gradient Algorithms}},}\
  }\href {https://doi.org/10.22331/q-2022-07-20-770} {\bibfield  {journal}
  {\bibinfo  {journal} {Quantum}\ }\textbf {\bibinfo {volume} {6}},\ \bibinfo
  {pages} {770} (\bibinfo {year} {2022})}\BibitemShut {NoStop}%
\bibitem [{\citenamefont {Markowitz}(1952)}]{Markowitz1952}%
  \BibitemOpen
  \bibfield  {author} {\bibinfo {author} {\bibfnamefont {H.}~\bibnamefont
  {Markowitz}},\ }\bibfield  {title} {\enquote {\bibinfo {title} {\emph
  {Portfolio Selection}},}\ }\href {https://doi.org/10.2307/2975974} {\bibfield
   {journal} {\bibinfo  {journal} {The Journal of Finance}\ }\textbf {\bibinfo
  {volume} {7}},\ \bibinfo {pages} {77} (\bibinfo {year} {1952})}\BibitemShut
  {NoStop}%
\bibitem [{\citenamefont {Roy}(1952)}]{Roy1952}%
  \BibitemOpen
  \bibfield  {author} {\bibinfo {author} {\bibfnamefont {A.~D.}\ \bibnamefont
  {Roy}},\ }\bibfield  {title} {\enquote {\bibinfo {title} {\emph {Safety First
  and the Holding of Assets}},}\ }\href {https://doi.org/10.2307/1907413}
  {\bibfield  {journal} {\bibinfo  {journal} {Econometrica}\ }\textbf {\bibinfo
  {volume} {20}},\ \bibinfo {pages} {431} (\bibinfo {year} {1952})}\BibitemShut
  {NoStop}%
\bibitem [{\citenamefont {Egger}\ \emph {et~al.}(2020)\citenamefont {Egger},
  \citenamefont {Gutierrez}, \citenamefont {Mestre},\ and\ \citenamefont
  {Woerner}}]{egger2019credit}%
  \BibitemOpen
  \bibfield  {author} {\bibinfo {author} {\bibfnamefont {D.~J.}\ \bibnamefont
  {Egger}}, \bibinfo {author} {\bibfnamefont {R.~G.}\ \bibnamefont
  {Gutierrez}}, \bibinfo {author} {\bibfnamefont {J.~C.}\ \bibnamefont
  {Mestre}}, \ and\ \bibinfo {author} {\bibfnamefont {S.}~\bibnamefont
  {Woerner}},\ }\bibfield  {title} {\enquote {\bibinfo {title} {\emph {Credit
  risk analysis using quantum computers}},}\ }\href
  {https://doi.org/10.1109/TC.2020.3038063} {\bibfield  {journal} {\bibinfo
  {journal} {IEEE Transactions on Computers}\ } (\bibinfo {year}
  {2020})}\BibitemShut {NoStop}%
\bibitem [{\citenamefont {Low}\ and\ \citenamefont
  {Chuang}(2017{\natexlab{a}})}]{Low2017optimal}%
  \BibitemOpen
  \bibfield  {author} {\bibinfo {author} {\bibfnamefont {G.~H.}\ \bibnamefont
  {Low}}\ and\ \bibinfo {author} {\bibfnamefont {I.~L.}\ \bibnamefont
  {Chuang}},\ }\bibfield  {title} {\enquote {\bibinfo {title} {\emph {Optimal
  Hamiltonian Simulation by Quantum Signal Processing}},}\ }\href
  {https://doi.org/10.1103/PhysRevLett.118.010501} {\bibfield  {journal}
  {\bibinfo  {journal} {Phys. Rev. Lett.}\ }\textbf {\bibinfo {volume} {118}},\
  \bibinfo {pages} {010501} (\bibinfo {year} {2017}{\natexlab{a}})}\BibitemShut
  {NoStop}%
\bibitem [{\citenamefont {Gily{\'e}n}\ \emph {et~al.}(2019)\citenamefont
  {Gily{\'e}n}, \citenamefont {Su}, \citenamefont {Low},\ and\ \citenamefont
  {Wiebe}}]{gilyen2019quantum}%
  \BibitemOpen
  \bibfield  {author} {\bibinfo {author} {\bibfnamefont {A.}~\bibnamefont
  {Gily{\'e}n}}, \bibinfo {author} {\bibfnamefont {Y.}~\bibnamefont {Su}},
  \bibinfo {author} {\bibfnamefont {G.~H.}\ \bibnamefont {Low}}, \ and\
  \bibinfo {author} {\bibfnamefont {N.}~\bibnamefont {Wiebe}},\ }\bibfield
  {title} {\enquote {\bibinfo {title} {\emph {Quantum singular value
  transformation and beyond: exponential improvements for quantum matrix
  arithmetics}},}\ }in\ \href {https://doi.org/10.1145/3313276.3316366}
  {\bibinfo {booktitle} {Proceedings of the 51st Annual ACM SIGACT Symposium on
  Theory of Computing}}\ (\bibinfo {year} {2019})\ pp.\ \bibinfo {pages}
  {193--204}\BibitemShut {NoStop}%
\bibitem [{\citenamefont {Low}\ and\ \citenamefont
  {Chuang}(2019)}]{low2019hamiltonian}%
  \BibitemOpen
  \bibfield  {author} {\bibinfo {author} {\bibfnamefont {G.~H.}\ \bibnamefont
  {Low}}\ and\ \bibinfo {author} {\bibfnamefont {I.~L.}\ \bibnamefont
  {Chuang}},\ }\bibfield  {title} {\enquote {\bibinfo {title} {\emph
  {Hamiltonian simulation by qubitization}},}\ }\href
  {https://doi.org/10.22331/q-2019-07-12-163} {\bibfield  {journal} {\bibinfo
  {journal} {Quantum}\ }\textbf {\bibinfo {volume} {3}},\ \bibinfo {pages}
  {163} (\bibinfo {year} {2019})}\BibitemShut {NoStop}%
\bibitem [{\citenamefont {Kikuchi}\ \emph {et~al.}(2023)\citenamefont
  {Kikuchi}, \citenamefont {Mc~Keever}, \citenamefont {Coopmans}, \citenamefont
  {Lubasch},\ and\ \citenamefont {Benedetti}}]{Kikuchi_2023}%
  \BibitemOpen
  \bibfield  {author} {\bibinfo {author} {\bibfnamefont {Y.}~\bibnamefont
  {Kikuchi}}, \bibinfo {author} {\bibfnamefont {C.}~\bibnamefont {Mc~Keever}},
  \bibinfo {author} {\bibfnamefont {L.}~\bibnamefont {Coopmans}}, \bibinfo
  {author} {\bibfnamefont {M.}~\bibnamefont {Lubasch}}, \ and\ \bibinfo
  {author} {\bibfnamefont {M.}~\bibnamefont {Benedetti}},\ }\bibfield  {title}
  {\enquote {\bibinfo {title} {\emph {Realization of quantum signal processing
  on a noisy quantum computer}},}\ }\href
  {https://doi.org/10.1038/s41534-023-00762-0} {\bibfield  {journal} {\bibinfo
  {journal} {npj Quantum Information}\ }\textbf {\bibinfo {volume} {9}}
  (\bibinfo {year} {2023})}\BibitemShut {NoStop}%
\bibitem [{\citenamefont {Martyn}\ \emph
  {et~al.}(2021{\natexlab{a}})\citenamefont {Martyn}, \citenamefont {Rossi},
  \citenamefont {Tan},\ and\ \citenamefont {Chuang}}]{martyn2021grand}%
  \BibitemOpen
  \bibfield  {author} {\bibinfo {author} {\bibfnamefont {J.~M.}\ \bibnamefont
  {Martyn}}, \bibinfo {author} {\bibfnamefont {Z.~M.}\ \bibnamefont {Rossi}},
  \bibinfo {author} {\bibfnamefont {A.~K.}\ \bibnamefont {Tan}}, \ and\
  \bibinfo {author} {\bibfnamefont {I.~L.}\ \bibnamefont {Chuang}},\ }\bibfield
   {title} {\enquote {\bibinfo {title} {\emph {Grand Unification of Quantum
  Algorithms}},}\ }\href {https://doi.org/10.1103/prxquantum.2.040203}
  {\bibfield  {journal} {\bibinfo  {journal} {PRX Quantum}\ }\textbf {\bibinfo
  {volume} {2}} (\bibinfo {year} {2021}{\natexlab{a}})}\BibitemShut {NoStop}%
\bibitem [{\citenamefont {Martyn}\ \emph
  {et~al.}(2021{\natexlab{b}})\citenamefont {Martyn}, \citenamefont {Liu},
  \citenamefont {Chin},\ and\ \citenamefont {Chuang}}]{martyn2021efficient}%
  \BibitemOpen
  \bibfield  {author} {\bibinfo {author} {\bibfnamefont {J.~M.}\ \bibnamefont
  {Martyn}}, \bibinfo {author} {\bibfnamefont {Y.}~\bibnamefont {Liu}},
  \bibinfo {author} {\bibfnamefont {Z.~E.}\ \bibnamefont {Chin}}, \ and\
  \bibinfo {author} {\bibfnamefont {I.~L.}\ \bibnamefont {Chuang}},\ }\bibfield
   {title} {\enquote {\bibinfo {title} {\emph {Efficient Fully-Coherent
  Hamiltonian Simulation}},}\ }\href@noop {} {\bibfield  {journal} {\bibinfo
  {journal} {arXiv preprint arXiv:2110.11327}\ } (\bibinfo {year}
  {2021}{\natexlab{b}})},\ \Eprint {http://arxiv.org/abs/2110.11327}
  {arXiv:2110.11327 [quant-ph]} \BibitemShut {NoStop}%
\bibitem [{\citenamefont {Lin}\ and\ \citenamefont {Tong}(2020)}]{Lin_2020}%
  \BibitemOpen
  \bibfield  {author} {\bibinfo {author} {\bibfnamefont {L.}~\bibnamefont
  {Lin}}\ and\ \bibinfo {author} {\bibfnamefont {Y.}~\bibnamefont {Tong}},\
  }\bibfield  {title} {\enquote {\bibinfo {title} {\emph {Optimal polynomial
  based quantum eigenstate filtering with application to solving quantum linear
  systems}},}\ }\href {https://doi.org/10.22331/q-2020-11-11-361} {\bibfield
  {journal} {\bibinfo  {journal} {Quantum}\ }\textbf {\bibinfo {volume} {4}},\
  \bibinfo {pages} {361} (\bibinfo {year} {2020})}\BibitemShut {NoStop}%
\bibitem [{\citenamefont {Rall}\ and\ \citenamefont
  {Fuller}(2023)}]{Rall_2023}%
  \BibitemOpen
  \bibfield  {author} {\bibinfo {author} {\bibfnamefont {P.}~\bibnamefont
  {Rall}}\ and\ \bibinfo {author} {\bibfnamefont {B.}~\bibnamefont {Fuller}},\
  }\bibfield  {title} {\enquote {\bibinfo {title} {\emph {Amplitude Estimation
  from Quantum Signal Processing}},}\ }\href
  {https://doi.org/10.22331/q-2023-03-02-937} {\bibfield  {journal} {\bibinfo
  {journal} {Quantum}\ }\textbf {\bibinfo {volume} {7}},\ \bibinfo {pages}
  {937} (\bibinfo {year} {2023})}\BibitemShut {NoStop}%
\bibitem [{\citenamefont {Stamatopoulos}\ and\ \citenamefont
  {Zeng}(2023)}]{stamatopoulos2023derivative}%
  \BibitemOpen
  \bibfield  {author} {\bibinfo {author} {\bibfnamefont {N.}~\bibnamefont
  {Stamatopoulos}}\ and\ \bibinfo {author} {\bibfnamefont {W.~J.}\ \bibnamefont
  {Zeng}},\ }\bibfield  {title} {\enquote {\bibinfo {title} {\emph {Derivative
  Pricing using Quantum Signal Processing}},}\ }\href@noop {} {\  (\bibinfo
  {year} {2023})},\ \Eprint {http://arxiv.org/abs/2307.14310} {arXiv:2307.14310
  [quant-ph]} \BibitemShut {NoStop}%
\bibitem [{\citenamefont {Gilyén}\ \emph {et~al.}(2019)\citenamefont
  {Gilyén}, \citenamefont {Arunachalam},\ and\ \citenamefont
  {Wiebe}}]{gilyen2019optimizing}%
  \BibitemOpen
  \bibfield  {author} {\bibinfo {author} {\bibfnamefont {A.}~\bibnamefont
  {Gilyén}}, \bibinfo {author} {\bibfnamefont {S.}~\bibnamefont
  {Arunachalam}}, \ and\ \bibinfo {author} {\bibfnamefont {N.}~\bibnamefont
  {Wiebe}},\ }\bibfield  {title} {\enquote {\bibinfo {title} {\emph {Optimizing
  quantum optimization algorithms via faster quantum gradient computation}},}\
  }\href {https://doi.org/10.1137/1.9781611975482.87} {\bibfield  {journal}
  {\bibinfo  {journal} {Proceedings of the Thirtieth Annual ACM-SIAM Symposium
  on Discrete Algorithms}\ ,\ \bibinfo {pages} {1425–1444}} (\bibinfo {year}
  {2019})}\BibitemShut {NoStop}%
\bibitem [{\citenamefont {Brassard}\ \emph {et~al.}(2002)\citenamefont
  {Brassard}, \citenamefont {Hoyer}, \citenamefont {Mosca},\ and\ \citenamefont
  {Tapp}}]{brassard2002quantum}%
  \BibitemOpen
  \bibfield  {author} {\bibinfo {author} {\bibfnamefont {G.}~\bibnamefont
  {Brassard}}, \bibinfo {author} {\bibfnamefont {P.}~\bibnamefont {Hoyer}},
  \bibinfo {author} {\bibfnamefont {M.}~\bibnamefont {Mosca}}, \ and\ \bibinfo
  {author} {\bibfnamefont {A.}~\bibnamefont {Tapp}},\ }\bibfield  {title}
  {\enquote {\bibinfo {title} {\emph {{Quantum Amplitude Amplification and
  Estimation}}},}\ }\href {https://doi.org/10.1090/conm/305/05215} {\bibfield
  {journal} {\bibinfo  {journal} {Contemporary Mathematics}\ }\textbf {\bibinfo
  {volume} {305}} (\bibinfo {year} {2002})}\BibitemShut {NoStop}%
\bibitem [{\citenamefont {Grinko}\ \emph {et~al.}(2021)\citenamefont {Grinko},
  \citenamefont {Gacon}, \citenamefont {Zoufal},\ and\ \citenamefont
  {Woerner}}]{grinko2021iterative}%
  \BibitemOpen
  \bibfield  {author} {\bibinfo {author} {\bibfnamefont {D.}~\bibnamefont
  {Grinko}}, \bibinfo {author} {\bibfnamefont {J.}~\bibnamefont {Gacon}},
  \bibinfo {author} {\bibfnamefont {C.}~\bibnamefont {Zoufal}}, \ and\ \bibinfo
  {author} {\bibfnamefont {S.}~\bibnamefont {Woerner}},\ }\bibfield  {title}
  {\enquote {\bibinfo {title} {\emph {Iterative quantum amplitude
  estimation}},}\ }\href {https://doi.org/10.1038/s41534-021-00379-1}
  {\bibfield  {journal} {\bibinfo  {journal} {npj Quantum Information}\
  }\textbf {\bibinfo {volume} {7}} (\bibinfo {year} {2021})}\BibitemShut
  {NoStop}%
\bibitem [{\citenamefont {Suzuki}\ \emph {et~al.}(2020)\citenamefont {Suzuki},
  \citenamefont {Uno}, \citenamefont {Raymond}, \citenamefont {Tanaka},
  \citenamefont {Onodera},\ and\ \citenamefont
  {Yamamoto}}]{suzuki2020amplitude}%
  \BibitemOpen
  \bibfield  {author} {\bibinfo {author} {\bibfnamefont {Y.}~\bibnamefont
  {Suzuki}}, \bibinfo {author} {\bibfnamefont {S.}~\bibnamefont {Uno}},
  \bibinfo {author} {\bibfnamefont {R.}~\bibnamefont {Raymond}}, \bibinfo
  {author} {\bibfnamefont {T.}~\bibnamefont {Tanaka}}, \bibinfo {author}
  {\bibfnamefont {T.}~\bibnamefont {Onodera}}, \ and\ \bibinfo {author}
  {\bibfnamefont {N.}~\bibnamefont {Yamamoto}},\ }\bibfield  {title} {\enquote
  {\bibinfo {title} {\emph {Amplitude estimation without phase estimation}},}\
  }\href {https://doi.org/10.1007/s11128-019-2565-2} {\bibfield  {journal}
  {\bibinfo  {journal} {Quantum Information Processing}\ }\textbf {\bibinfo
  {volume} {19}},\ \bibinfo {pages} {75} (\bibinfo {year} {2020})}\BibitemShut
  {NoStop}%
\bibitem [{\citenamefont {Giurgica-Tiron}\ \emph {et~al.}(2022)\citenamefont
  {Giurgica-Tiron}, \citenamefont {Kerenidis}, \citenamefont {Labib},
  \citenamefont {Prakash},\ and\ \citenamefont {Zeng}}]{Giurgica_Tiron_2022}%
  \BibitemOpen
  \bibfield  {author} {\bibinfo {author} {\bibfnamefont {T.}~\bibnamefont
  {Giurgica-Tiron}}, \bibinfo {author} {\bibfnamefont {I.}~\bibnamefont
  {Kerenidis}}, \bibinfo {author} {\bibfnamefont {F.}~\bibnamefont {Labib}},
  \bibinfo {author} {\bibfnamefont {A.}~\bibnamefont {Prakash}}, \ and\
  \bibinfo {author} {\bibfnamefont {W.}~\bibnamefont {Zeng}},\ }\bibfield
  {title} {\enquote {\bibinfo {title} {\emph {Low depth algorithms for quantum
  amplitude estimation}},}\ }\href {https://doi.org/10.22331/q-2022-06-27-745}
  {\bibfield  {journal} {\bibinfo  {journal} {Quantum}\ }\textbf {\bibinfo
  {volume} {6}},\ \bibinfo {pages} {745} (\bibinfo {year} {2022})}\BibitemShut
  {NoStop}%
\bibitem [{\citenamefont {Rall}(2021)}]{Rall_2021}%
  \BibitemOpen
  \bibfield  {author} {\bibinfo {author} {\bibfnamefont {P.}~\bibnamefont
  {Rall}},\ }\bibfield  {title} {\enquote {\bibinfo {title} {\emph {Faster
  Coherent Quantum Algorithms for Phase, Energy, and Amplitude Estimation}},}\
  }\href {https://doi.org/10.22331/q-2021-10-19-566} {\bibfield  {journal}
  {\bibinfo  {journal} {Quantum}\ }\textbf {\bibinfo {volume} {5}},\ \bibinfo
  {pages} {566} (\bibinfo {year} {2021})}\BibitemShut {NoStop}%
\bibitem [{\citenamefont {Haah}(2019)}]{Haah2019product}%
  \BibitemOpen
  \bibfield  {author} {\bibinfo {author} {\bibfnamefont {J.}~\bibnamefont
  {Haah}},\ }\bibfield  {title} {\enquote {\bibinfo {title} {\emph {Product
  {D}ecomposition of {P}eriodic {F}unctions in {Q}uantum {S}ignal
  {P}rocessing}},}\ }\href {https://doi.org/10.22331/q-2019-10-07-190}
  {\bibfield  {journal} {\bibinfo  {journal} {{Quantum}}\ }\textbf {\bibinfo
  {volume} {3}},\ \bibinfo {pages} {190} (\bibinfo {year} {2019})}\BibitemShut
  {NoStop}%
\bibitem [{\citenamefont {Chao}\ \emph {et~al.}(2020)\citenamefont {Chao},
  \citenamefont {Ding}, \citenamefont {Gilyen}, \citenamefont {Huang},\ and\
  \citenamefont {Szegedy}}]{chao2020finding}%
  \BibitemOpen
  \bibfield  {author} {\bibinfo {author} {\bibfnamefont {R.}~\bibnamefont
  {Chao}}, \bibinfo {author} {\bibfnamefont {D.}~\bibnamefont {Ding}}, \bibinfo
  {author} {\bibfnamefont {A.}~\bibnamefont {Gilyen}}, \bibinfo {author}
  {\bibfnamefont {C.}~\bibnamefont {Huang}}, \ and\ \bibinfo {author}
  {\bibfnamefont {M.}~\bibnamefont {Szegedy}},\ }\bibfield  {title} {\enquote
  {\bibinfo {title} {\emph {Finding Angles for Quantum Signal Processing with
  Machine Precision}},}\ }\href@noop {} {\bibfield  {journal} {\bibinfo
  {journal} {arXiv preprint arXiv:2003.02831}\ } (\bibinfo {year} {2020})},\
  \Eprint {http://arxiv.org/abs/2003.02831} {arXiv:2003.02831 [quant-ph]}
  \BibitemShut {NoStop}%
\bibitem [{\citenamefont {Dong}\ \emph {et~al.}(2021)\citenamefont {Dong},
  \citenamefont {Meng}, \citenamefont {Whaley},\ and\ \citenamefont
  {Lin}}]{dong2021efficient}%
  \BibitemOpen
  \bibfield  {author} {\bibinfo {author} {\bibfnamefont {Y.}~\bibnamefont
  {Dong}}, \bibinfo {author} {\bibfnamefont {X.}~\bibnamefont {Meng}}, \bibinfo
  {author} {\bibfnamefont {K.~B.}\ \bibnamefont {Whaley}}, \ and\ \bibinfo
  {author} {\bibfnamefont {L.}~\bibnamefont {Lin}},\ }\bibfield  {title}
  {\enquote {\bibinfo {title} {\emph {Efficient phase-factor evaluation in
  quantum signal processing}},}\ }\href
  {https://doi.org/10.1103/physreva.103.042419} {\bibfield  {journal} {\bibinfo
   {journal} {Physical Review A}\ }\textbf {\bibinfo {volume} {103}},\ \bibinfo
  {pages} {042419} (\bibinfo {year} {2021})}\BibitemShut {NoStop}%
\bibitem [{\citenamefont {Dong}\ \emph {et~al.}(2022)\citenamefont {Dong},
  \citenamefont {Lin},\ and\ \citenamefont {Tong}}]{dong2022ground}%
  \BibitemOpen
  \bibfield  {author} {\bibinfo {author} {\bibfnamefont {Y.}~\bibnamefont
  {Dong}}, \bibinfo {author} {\bibfnamefont {L.}~\bibnamefont {Lin}}, \ and\
  \bibinfo {author} {\bibfnamefont {Y.}~\bibnamefont {Tong}},\ }\bibfield
  {title} {\enquote {\bibinfo {title} {\emph {Ground-State Preparation and
  Energy Estimation on Early Fault-Tolerant Quantum Computers via Quantum
  Eigenvalue Transformation of Unitary Matrices}},}\ }\href
  {https://doi.org/10.1103/prxquantum.3.040305} {\bibfield  {journal} {\bibinfo
   {journal} {{PRX} Quantum}\ }\textbf {\bibinfo {volume} {3}} (\bibinfo {year}
  {2022})}\BibitemShut {NoStop}%
\bibitem [{\citenamefont {Low}\ and\ \citenamefont
  {Chuang}(2017{\natexlab{b}})}]{low2017hamiltonian}%
  \BibitemOpen
  \bibfield  {author} {\bibinfo {author} {\bibfnamefont {G.~H.}\ \bibnamefont
  {Low}}\ and\ \bibinfo {author} {\bibfnamefont {I.~L.}\ \bibnamefont
  {Chuang}},\ }\bibfield  {title} {\enquote {\bibinfo {title} {\emph
  {Hamiltonian simulation by uniform spectral amplification}},}\ }\href
  {https://doi.org/10.48550/ARXIV.1707.05391} {\bibfield  {journal} {\bibinfo
  {journal} {arXiv preprint arXiv:1707.05391}\ } (\bibinfo {year}
  {2017}{\natexlab{b}})}\BibitemShut {NoStop}%
\bibitem [{\citenamefont {Clader}\ \emph {et~al.}(2022)\citenamefont {Clader},
  \citenamefont {Dalzell}, \citenamefont {Stamatopoulos}, \citenamefont
  {Salton}, \citenamefont {Berta},\ and\ \citenamefont
  {Zeng}}]{clader2022quantum}%
  \BibitemOpen
  \bibfield  {author} {\bibinfo {author} {\bibfnamefont {B.~D.}\ \bibnamefont
  {Clader}}, \bibinfo {author} {\bibfnamefont {A.~M.}\ \bibnamefont {Dalzell}},
  \bibinfo {author} {\bibfnamefont {N.}~\bibnamefont {Stamatopoulos}}, \bibinfo
  {author} {\bibfnamefont {G.}~\bibnamefont {Salton}}, \bibinfo {author}
  {\bibfnamefont {M.}~\bibnamefont {Berta}}, \ and\ \bibinfo {author}
  {\bibfnamefont {W.~J.}\ \bibnamefont {Zeng}},\ }\bibfield  {title} {\enquote
  {\bibinfo {title} {\emph {Quantum Resources Required to Block-Encode a Matrix
  of Classical Data}},}\ }\href {https://doi.org/10.1109/tqe.2022.3231194}
  {\bibfield  {journal} {\bibinfo  {journal} {IEEE Transactions on Quantum
  Engineering}\ }\textbf {\bibinfo {volume} {3}},\ \bibinfo {pages} {1–23}
  (\bibinfo {year} {2022})}\BibitemShut {NoStop}%
\bibitem [{\citenamefont {O’Brien}\ and\ \citenamefont
  {Szerszeń}(2017)}]{OBRIEN2017215}%
  \BibitemOpen
  \bibfield  {author} {\bibinfo {author} {\bibfnamefont {J.}~\bibnamefont
  {O’Brien}}\ and\ \bibinfo {author} {\bibfnamefont {P.~J.}\ \bibnamefont
  {Szerszeń}},\ }\bibfield  {title} {\enquote {\bibinfo {title} {\emph {An
  evaluation of bank measures for market risk before, during and after the
  financial crisis}},}\ }\href
  {https://doi.org/https://doi.org/10.1016/j.jbankfin.2017.03.002} {\bibfield
  {journal} {\bibinfo  {journal} {Journal of Banking \& Finance}\ }\textbf
  {\bibinfo {volume} {80}},\ \bibinfo {pages} {215} (\bibinfo {year}
  {2017})}\BibitemShut {NoStop}%
\bibitem [{\citenamefont {Milek}(2020)}]{milek2020quantum}%
  \BibitemOpen
  \bibfield  {author} {\bibinfo {author} {\bibfnamefont {J.}~\bibnamefont
  {Milek}},\ }\bibfield  {title} {\enquote {\bibinfo {title} {\emph {Quantum
  Implementation of Risk Analysis-relevant Copulas}},}\ }\href
  {https://doi.org/10.48550/ARXIV.2002.07389} {\bibfield  {journal} {\bibinfo
  {journal} {arXiv preprint arXiv:2002.07389}\ } (\bibinfo {year}
  {2020})}\BibitemShut {NoStop}%
\bibitem [{\citenamefont {Zhu}\ \emph {et~al.}(2022)\citenamefont {Zhu},
  \citenamefont {Shen}, \citenamefont {Giani}, \citenamefont {Majumder},
  \citenamefont {Neculaes},\ and\ \citenamefont {Johri}}]{zhu2022copula}%
  \BibitemOpen
  \bibfield  {author} {\bibinfo {author} {\bibfnamefont {D.}~\bibnamefont
  {Zhu}}, \bibinfo {author} {\bibfnamefont {W.}~\bibnamefont {Shen}}, \bibinfo
  {author} {\bibfnamefont {A.}~\bibnamefont {Giani}}, \bibinfo {author}
  {\bibfnamefont {S.~R.}\ \bibnamefont {Majumder}}, \bibinfo {author}
  {\bibfnamefont {B.}~\bibnamefont {Neculaes}}, \ and\ \bibinfo {author}
  {\bibfnamefont {S.}~\bibnamefont {Johri}},\ }\bibfield  {title} {\enquote
  {\bibinfo {title} {\emph {Copula-based Risk Aggregation with Trapped Ion
  Quantum Computers}},}\ }\href {https://doi.org/10.48550/ARXIV.2206.11937}
  {\bibfield  {journal} {\bibinfo  {journal} {arXiv preprint arXiv:2206.11937}\
  } (\bibinfo {year} {2022})}\BibitemShut {NoStop}%
\bibitem [{\citenamefont {Ross}\ and\ \citenamefont
  {Selinger}(2016)}]{ross_2016}%
  \BibitemOpen
  \bibfield  {author} {\bibinfo {author} {\bibfnamefont {N.~J.}\ \bibnamefont
  {Ross}}\ and\ \bibinfo {author} {\bibfnamefont {P.}~\bibnamefont
  {Selinger}},\ }\bibfield  {title} {\enquote {\bibinfo {title} {\emph {Optimal
  Ancilla-Free Clifford+T Approximation of z-Rotations}},}\ }\href
  {https://dl.acm.org/doi/abs/10.5555/3179330.3179331} {\bibfield  {journal}
  {\bibinfo  {journal} {Quantum Info. Comput.}\ }\textbf {\bibinfo {volume}
  {16}},\ \bibinfo {pages} {901} (\bibinfo {year} {2016})}\BibitemShut
  {NoStop}%
\bibitem [{\citenamefont {Kim}\ and\ \citenamefont
  {Choi}(2018)}]{Kim_Efficient_2018}%
  \BibitemOpen
  \bibfield  {author} {\bibinfo {author} {\bibfnamefont {T.}~\bibnamefont
  {Kim}}\ and\ \bibinfo {author} {\bibfnamefont {B.}~\bibnamefont {Choi}},\
  }\bibfield  {title} {\enquote {\bibinfo {title} {\emph {Efficient
  decomposition methods for controlled-R n using a single ancillary qubit}},}\
  }\href {https://doi.org/10.1038/s41598-018-23764-x} {\bibfield  {journal}
  {\bibinfo  {journal} {Scientific Reports}\ }\textbf {\bibinfo {volume} {8}}
  (\bibinfo {year} {2018})}\BibitemShut {NoStop}%
\bibitem [{\citenamefont {Shukla}\ and\ \citenamefont
  {Vedula}(2024)}]{Shukla_2024}%
  \BibitemOpen
  \bibfield  {author} {\bibinfo {author} {\bibfnamefont {A.}~\bibnamefont
  {Shukla}}\ and\ \bibinfo {author} {\bibfnamefont {P.}~\bibnamefont
  {Vedula}},\ }\bibfield  {title} {\enquote {\bibinfo {title} {\emph {An
  efficient quantum algorithm for preparation of uniform quantum superposition
  states}},}\ }\href {https://doi.org/10.1007/s11128-024-04258-4} {\bibfield
  {journal} {\bibinfo  {journal} {Quantum Information Processing}\ }\textbf
  {\bibinfo {volume} {23}} (\bibinfo {year} {2024})}\BibitemShut {NoStop}%
\bibitem [{\citenamefont {Giles}\ and\ \citenamefont
  {Haji-Ali}(2019)}]{giles2019}%
  \BibitemOpen
  \bibfield  {author} {\bibinfo {author} {\bibfnamefont {M.~B.}\ \bibnamefont
  {Giles}}\ and\ \bibinfo {author} {\bibfnamefont {A.-L.}\ \bibnamefont
  {Haji-Ali}},\ }\bibfield  {title} {\enquote {\bibinfo {title} {\emph
  {Multilevel Nested Simulation for Efficient Risk Estimation}},}\ }\href
  {https://doi.org/10.1137/18M1173186} {\bibfield  {journal} {\bibinfo
  {journal} {SIAM/ASA Journal on Uncertainty Quantification}\ }\textbf
  {\bibinfo {volume} {7}},\ \bibinfo {pages} {497} (\bibinfo {year}
  {2019})}\BibitemShut {NoStop}%
\bibitem [{\citenamefont {An}\ \emph {et~al.}(2021)\citenamefont {An},
  \citenamefont {Linden}, \citenamefont {Liu}, \citenamefont {Montanaro},
  \citenamefont {Shao},\ and\ \citenamefont {Wang}}]{An_2021}%
  \BibitemOpen
  \bibfield  {author} {\bibinfo {author} {\bibfnamefont {D.}~\bibnamefont
  {An}}, \bibinfo {author} {\bibfnamefont {N.}~\bibnamefont {Linden}}, \bibinfo
  {author} {\bibfnamefont {J.-P.}\ \bibnamefont {Liu}}, \bibinfo {author}
  {\bibfnamefont {A.}~\bibnamefont {Montanaro}}, \bibinfo {author}
  {\bibfnamefont {C.}~\bibnamefont {Shao}}, \ and\ \bibinfo {author}
  {\bibfnamefont {J.}~\bibnamefont {Wang}},\ }\bibfield  {title} {\enquote
  {\bibinfo {title} {\emph {Quantum-accelerated multilevel Monte Carlo methods
  for stochastic differential equations in mathematical finance}},}\ }\href
  {https://doi.org/10.22331/q-2021-06-24-481} {\bibfield  {journal} {\bibinfo
  {journal} {Quantum}\ }\textbf {\bibinfo {volume} {5}},\ \bibinfo {pages}
  {481} (\bibinfo {year} {2021})}\BibitemShut {NoStop}%
\end{thebibliography}%

\appendix

\section{Derivative Pricing using Amplitude Estimation}
\label{app:ae}
The price of a derivative contract is calculated as the expectation value of the derivative's discounted payoff, evaluated over a suitable stochastic process that is assumed to govern the dynamics of the underlying market parameters.
Given stochastic paths $\omega \in \Omega$, each of which occuring with probability $p(\omega)$, the expectation value of the discounted payoff $f(\omega)$ can be written as

\begin{equation}
	\label{eqn:derivative_exp_value}
\mathbb{E}[f] = \sum_{\omega \in \Omega}p(\omega)f(\omega).
\end{equation}

Quantum Amplitude Estimation (QAE) \cite{brassard2002quantum, montanaro2015quantum} can be used to estimate such expectation values if the quantity of interest can be efficiently encoded as the probability of a specific measurement outcome.
QAE takes as input a unitary operator $\mathcal{A}$ which produces the state
\begin{equation}
	\label{eqn:A_operator_app}
    \mathcal{A} : \ket{0}_{n+1} \rightarrow \left( \sqrt{V}\ket{\psi_0}_n\ket{0} + \sqrt{1-V}\ket{\psi_1}_n\ket{1} \right),
\end{equation}
where $\ket{\psi_0}$ and $\ket{\psi_1}$ are arbitrary, normalized quantum states and estimates the derivative price $V$ to additive error $\epsilon$ using $\bigO{1/\epsilon}$ invocations of the operator $\mathcal{Q} = \mathcal{A}S_0\mathcal{A}^{\dagger}S _{\psi_1}$, where $S_0 = \mathbb{I} - 2 |0\rangle_{n+1}\langle 0|_{n+1}$ and $S_{\psi_1} = \mathbb{I} - 2\ket{1}\ket{\psi_1}_n\bra{1}\bra{\psi_1}_n$.

A method to construct the $\mathcal{A}$ operator of Eq.~\eqref{eqn:A_operator_app} for typical derivative contracts of practical interest and the corresponding required quantum resources is shown in \cite{chakrabarti2021threshold}.
For a contract with $d$ stochastic market parameters $X = [x_1, x_2, \cdots x_d]$, an operator $\mathcal{P}$ constructs a probability-weighted superposition of all possible combinations of stochastic realizations of the market parameters,

\begin{equation}
	\label{eqn:derivative_path_loading}
	\mathcal{P} :\ket{\vec{0}}^{\otimes d} \rightarrow \sum_{\omega}\sqrt{p(\omega)}\ket{\omega},
\end{equation}
with an appropriate discretization for each parameter chosen such that the overall error from such discretization satisfies a desired accuracy.
Typically, each market parameter is modeled stochastically at pre-defined points in time that are determined by the definition of the derivative contract.
As such, each $\ket{\omega}$ contains all the information about the values of the market parameters $X$ for each \emph{path} of the stochastic process.
Then, an operator $\mathcal{F}$ computes the discounted payoff $f$ of the derivative on each path $\ket{\omega}$, and encodes that value in the amplitude of an ancilla qubit

\begin{equation}
	\label{eqn:derivative_payoff}
	\mathcal{F}:\ket{\omega}\ket{\vec{0}}\ket{0} \rightarrow \ket{\omega}\ket{f(\omega)} \left( \sqrt{f(\omega)}\ket{0} + \sqrt{1-f(\omega)}\ket{1} \right).
\end{equation}
Taking $\mathcal{A} = \mathcal{F}\mathcal{P}$

\begin{equation}
	\mathcal{F}\mathcal{P} : \ket{\vec{0}}^{\otimes d}\ket{0} \rightarrow \left( \sum_{\omega}\sqrt{p(\omega)}\sqrt{f(\omega)}\ket{\omega}\ket{f(\omega)}\ket{0} +  \sum_{\omega}\sqrt{p(\omega)}\sqrt{1-f(\omega)}\ket{\omega}\ket{f(\omega)}\ket{1} \right),
\end{equation}
gives us the desired operator of Eq.~\eqref{eqn:A_operator_app}, noticing that the probability of measuring $\ket{0}$ in the last qubit is the expectation value of the discounted payoff, as defined in Eq.~\eqref{eqn:derivative_exp_value}.

This approach can be generalized to compute the value $V$ of a derivative portfolio consisting of $m$ derivatives.
The $\mathcal{P}$ operator of Eq.~\eqref{eqn:derivative_path_loading} in this case will construct the stochastic process of the $d$ market parameters of the entire portfolio, and the $\mathcal{F}$ operator of Eq.~\eqref{eqn:derivative_payoff} will compute the sum of the payoffs of each derivative

\begin{equation}
	\label{eqn:portfolio_payoff}
	\mathcal{F}:\ket{\omega}\ket{\vec{0}}\ket{0} \rightarrow \ket{\omega}\ket{\sum_{i=1}^mf_i(\omega)} \left( \sqrt{\sum_{i=1}^mf_i(\omega)}\ket{0} + \sqrt{1-\sum_{i=1}^mf_i(\omega)}\ket{1} \right),
\end{equation}
where $f_i$ is the payoff of derivative $i \in [1,m]$.
Similarly to the single-derivative case, QAE can then be used to estimate the expectation value $V$ of the portfolio.

\begin{equation}
	V = \sum_{\omega \in \Omega}p(\omega)\sum_{i=1}^mf_i(\omega).
\end{equation}

\end{document}